%% file: 2nd_MAGIC_EHBL_catalog.tex
\documentclass[twocolumn]{aastex62}  

\graphicspath{{./}{}}
\newcommand\lf[1]{{#1}} 

\shorttitle{$2^{\text{nd}}$ MAGIC catalog of extreme blazars}
\shortauthors{$2^{\text{nd}}$ MAGIC catalog of extreme blazars}

\usepackage{natbib}
\usepackage{multirow}
\usepackage{threeparttable}
\usepackage[capitalise]{cleveref}
\usepackage{soul}
\usepackage{footmisc}
\usepackage{comment}
\usepackage[caption=false]{subfig}	

\NewPageAfterKeywords

\begin{document}

\title{Extreme Blazars Observed with MAGIC: Second Catalog Release}

\input{authorlist.tex}

\correspondingauthor{L.~Foffano, C.~Arcaro}
\email{contact.magic@mpp.mpg.de, lst-contact@cta-observatory.org}

\begin{abstract}
Extremely high-peaked BL Lac objects - also named \textit{extreme blazars} - are among the most energetic and persistent extragalactic accelerators in the Universe, defined by a synchrotron emission peaking above $10^{17}$ Hz in X-rays. Such emission is then reprocessed and produces radiation extending deeply into very-high-energy (VHE, energy E$>$100 GeV) gamma rays. Observations in this energy band - optimally investigated by the {Imaging Air-Shower Cherenkov} telescopes -  are crucial for probing the physical processes that drive their extreme behavior. \\
This study extends our investigation of extreme blazars in the VHE gamma-ray range, providing a second new mini-catalog of sources observed by the MAGIC telescopes. 
We report on the monitoring of seven targets between 2017 and 2025, including four newly observed sources and three that have been part of long-term observation campaigns, for a total of approximately 338 hours of observations.
The analysis of MAGIC data reveals two new VHE detections of extreme blazars, along with three additional sources showing hints of VHE emission.
Joint observations of MAGIC and the first Large-Sized Telescope (LST-1) also confirmed a new VHE extreme blazar. 
Our results are complemented by simultaneous multiwavelength observations in other energy bands, including optical-UV, X-rays, and  high-energy gamma rays (100 MeV$<$E$<$100 GeV). 
We confirm typical behavior of extreme blazars, such as a modest variability and a ``harder-when-brighter'' trend in X-rays across the sample.
This new set increases the population of extreme blazars and their broadband analysis confirms the physical properties of these extreme sources. 
\end{abstract}


\section{Introduction} 
\label{sec:intro}

\noindent
The nuclei of numerous galaxies host a supermassive black hole ($>$10$^6$ solar masses). These black holes accrete matter, and in a fraction of cases, the infalling energy is also re-emitted in the form of highly collimated jets emerging from the central region.  Such systems are referred to as active galactic nuclei (AGNs). When the jet axis is oriented close to the observer’s line of sight, the sources are classified as \textit{blazars}.

The radiation emitted by these objects ranges over the whole electromagnetic spectrum from radio to gamma rays, and is often influenced by relativistic effects. Its origin is mostly nonthermal, and it is usually dominated by synchrotron emission of relativistic accelerated electrons, which extends from radio up to X-ray energies. A second emission component, extending from X-rays to the deep gamma-ray range,  is often observed. In leptonic models, this component arises from the reprocessing of low-energy photons, although alternative processes may also contribute.
The spectral energy distribution (SED) of blazars is then usually characterized by these two components, which compose its typical two-bump structure.

Blazars are composed by a wide population of sources, which are sub-classified mostly into flat spectrum radio quasars (FSRQs) and BL Lac objects (BL Lacs), depending on the equivalent widths of emission lines in the optical spectrum \citep[e.g.][]{Urry95, Falomo2014}. The origin of this difference is likely related both to the different environmental conditions and to the evolutionary stage of these sources \citep{Cavaliere:2002}.

{BL Lac objects are further subdivided} depending on the energetics of the spectral emission peaks, and more precisely on the synchrotron-peak frequency $\nu_{\text{peak}}^{\text{sync}}$. 
Among these subcategories, there is an inverse relation between the observed bolometric luminosity and the maximum energy of the emission peaks, which has received different interpretations \citep{Fossati98, blazarsequence08, blazarsequence17}.
In this paper, we will discuss the two most energetic groups of sources, namely high-peaked BL Lac objects (HBL, $\nu_{\text{peak}}^{\text{sync}}$ between $10^{15}$ and $10^{17}$ Hz, adopting the definition of \citealt{2010ApJ...716...30A}) and extremely high-peaked objects (EHBLs or also \emph{extreme blazars}, $\nu_{\text{peak}}^{\text{sync}} > 10^{17}$~Hz, approximately 0.3~keV).

The extreme blazar class was first identified by \citet{2001AA...371..512C}.   
They are characterized by a synchrotron peak in the hard X-ray band and a high-energy peak which rises through the high-energy (HE, 100 MeV$<$E$<$100 GeV) band and peaks in the very-high-energy (VHE, E$>$100 GeV) gamma-ray band.
With respect to other blazars, they are characterized by lower luminosity, which affects their detectability in several energy bands. Due to the shift of the peaks to the highest energies, their HE gamma-ray emission is usually quite faint \citep[e.g.,][]{Paliya2019}. The typical flux variability in this energy band is also rather low, even if the current results are probably biased by the sensitivity limits of the \textit{Fermi} Large Area Telescope \citep[LAT,][]{2009ApJ...697.1071A}, mostly adopted for the investigation in this band \citep[see e.g.,][]{Nievas2022, lainez2025}.
In the VHE gamma-ray band, the high-energy bump shows the flux peak. Despite their  luminosity is lower with respect to other blazars, some objects can be investigated at these energies with the imaging atmospheric Cherenkov telescopes (IACTs), such as MAGIC and the first Large-Sized Telescope (LST-1). When selecting the most promising TeV-emitting candidates, it is important to keep in mind that this energy range is affected by the interactions with the extragalactic background Light \citep[EBL, see e.g.][]{Hauser:2001, Franceschini08, Franceschini17, Dominguez:2010bv}. When VHE gamma-ray photons interact with the EBL, producing electron–positron pairs, the observable flux in this band consequently, decreases. The impact of this attenuation increases with both the distance of the source and the energy of the gamma-ray photons, making distant sources difficult to detect at TeV energies.

During the past two decades, EHBLs have emerged as an intriguing blazar subclass, attracting increasing interest from both experimental and theoretical perspectives \citep[e.g.,][]{2020NatAs...4..124B}. Experimentally, several new sources have been detected and characterized at critical energies. As the EHBL population grows, spectral differences are emerging, particularly at the highest energies \citep{foffano2018}. The earliest identified EHBLs, like the archetypal 1ES~0229+200, now define only a specific subclass known as \emph{hard-TeV} sources, which display a very hard and stable gamma-ray spectrum peaking above 10 TeV. 
In contrast, newly discovered EHBLs with less extreme spectral features likely represent a transitional stage between HBLs and EHBLs, referred to as \emph{soft-TeV} sources \citep[e.g.,][]{PGC_MAGIC_paper}. These sources experience moderate flux variations, often showing temporarily characteristics of EHBLs during flaring episodes, as for example the case of 1ES~2344+514 \citep{2020A&A...640A.132M, 2024A&A...682A.114M} and 1ES~1959+650 \citep{2020A&A...638A..14M}. These differences show how VHE gamma-ray observations are essential for distinguishing these objects and understanding the underlying physics behind this blazar-class transition.

From a theoretical perspective, the emerging new sources have further challenged our {comprehension} of these objects and of the transitional behavior between classes. \emph{Soft-TeV} sources can often be well described by standard one-zone leptonic {scenarios}, such as the Synchrotron Self-Compton model \citep[SSC,][]{Maraschi:SSC, Tavecchio:SSC}. However, \emph{hard-TeV} objects with stable emission above several TeV are challenging to explain within this framework alone, as their SEDs require very extreme physical parameters. These models require  high minimum-electron Lorentz factors \citep{Lefa2011} and low magnetic fields in the emission region \citep{Tavecchio2011, Costamante2018}, taking the jet energetics far from equipartition between kinetic and magnetic energy \citep{tavecchio09}. 
These physical parameters support a scenario where the characteristically hard SEDs of \emph{hard-TeV} are produced by efficient particle acceleration and minimal radiative losses. 
More complex approaches, such as multizone or lepto-hadronic models \citep{Costamante2018}, are often needed for \emph{hard-TeV} EHBLs, incorporating multiple particle types and processes, such as shocks or magnetic reconnection \citep{sironi2015}. Nevertheless, some of these models, like the proton-synchrotron scenario \citep{cerruti2015}, struggle to reach the highest observed energies {and require} high jet power. Recently, models proposing complex re-acceleration mechanisms involving multiple shocks have shown to be promising in explaining the SEDs of some EHBLs \citep{Zech2021}.

The steadily growing number of EHBLs in the last 25 years has provided new insights into their nature and their connection to other known blazar subclasses. The aim of this paper is to investigate newly identified members of the EHBL population at TeV gamma-ray energies, presenting the most up-to-date results from the long-term observational program carried out by the MAGIC Telescopes Collaboration in the search for new TeV-emitting extreme blazars. \\

\noindent
The paper is structured as follows. In Section~\ref{sec:sources}, we describe the seven targets presented in this work. Sections~\ref{sec:MAGIC_results},~\ref{sec:Fermi_results},~\ref{sec:Swift_results}, and \ref{sec:tuorla_results} report the results of MAGIC, LST-1, {\it Fermi}-LAT, {\it Swift}-XRT/UVOT, and optical observations from the Tuorla observatory. 
The multiwavelength SED data and models are reported and discussed in Section~\ref{sec:modelling}. The details of the data analyses in the various bands as well as those of the modeling are reported in the Appendices.

\section{Sample description}
\label{sec:sources}
\noindent
The source selection is based on a set of phenomenological criteria, which can be summarized as follows: 
\begin{description}
    \item[X-ray behavior] EHBLs are supposed to be relatively well detected in X-rays as the synchrotron peak emission occurs in this energy range. Their {flux is usually rather high ($>10^{-12}$ erg cm$^{-2}$ s$^{-1}$}) and the  photon index is often relatively hard {($<$2)} in the 1-10 keV energy range. 
    \item[HE gamma-ray behavior] EHBLs are expected to be relatively faint in the HE gamma-ray range because of the spectral transition between the synchrotron and the inverse Compton (IC) emission, with rather hard photon indices due to the rising part of the high-energy spectral component.
    As a consequence, the signal detection of these sources usually requires long integration time {(months to years)}.
    \item[X-ray-to-radio and X-ray-to-gamma flux ratios] Several sources have shown higher chances to be detectable at VHE gamma rays when characterized by high X-ray-to-radio \citep[$>10^4$, see][]{bonnoli2015} and X-ray-to-gamma flux ratios \citep[$>1.15$, see][]{2020MNRAS.491.2771C}.
    \item[Multiwavelength correlation studies] Other studies have been conducted on the properties multiwavelength (MWL) of EHBLs, extracting TeV candidates based on correlations between different energy bands \citep{vandad_2017}.
\end{description}
Among the wide set of sources resulting from the previous criteria, we have chosen a set of seven interesting EHBLs to be observed -- also in the form of long-term monitoring -- by the MAGIC telescopes. 
Here, we present a summary of each source. All technical details such as coordinates, redshift, and historical estimates of the synchrotron peak extracted from the 2WHSP catalog \citep{Chang17} are reported in  \Cref{tab:source_list}. Flux and spectral indices comparisons in the X-ray and HE gamma-ray bands are also discussed in the next sections. 
\begin{description}
    \item[RBS~42] This source is located at a redshift of $z = 0.1$ \citep{ROSAT1998AN....319..347F} or larger \citep{2013ApJ...764..135S, 1996A&A...309..419N}. Its selection was based on the recent work of \citet{2020MNRAS.491.2771C}, where a list of EHBL candidates is proposed according to a high X-ray-to-gamma-ray flux ratio, in this case of 1.19. This source -- among those candidates with redshift below 0.3 -- shows the highest gamma-ray flux at 1\,GeV. 
    \item[TXS~0637--128] Located at moderately low redshift $z = 0.137$, recently determined by \citet{2020MNRAS.497...94P}, this BL Lac object was selected from the third catalog of hard \textit{Fermi}-LAT sources (3FHL catalog, \citealt{fermi3FHL}), where it is reported as an AGN of unknown type and with a hard photon index {$\Gamma _{\mathrm{3FHL}} = 1.63\pm 0.19$.} The source shows a high synchrotron peak at $10^{17.4}$~Hz and a high X-ray {flux F$_{(2-10\,\text{keV})} = (35.8 \pm 6.9) \times $10$^{-12}$ erg cm$^{-2}$ s$^{-1}$. }
    Observed by MAGIC in 2017 reporting a significance of $1.7\sigma$ over 12~h of good-quality data \citep{1st_MAGIC_EHBL_catalog}, it has been re-observed with a refined observation policy aimed at improving its detectability, lowering the energy threshold with dark-only observations of the source.
    \item[RX~J0805.4+7534] Selected as an interesting candidate among the list provided by \citet{vandad_2017} adopting MWL correlations, this source is located at moderate redshift $z = 0.121$ \citep{1996A&A...309..419N}.  Among all our candidates, it shows a high flux both in X-rays F$_{(2-10\,\text{keV})} = (56.2 \pm 35.2) \times $10$^{-12}$ erg cm$^{-2}$ s$^{-1}$ and    HE gamma~rays $F_{\mathrm{4FGL}} = (14.53 \pm 0.47) \times 10^{-10}$ ph cm$^{-2}$ s$^{-1}$. In the fourth \textit{Fermi}-LAT Catalog of gamma-ray sources (4FGL, \citealt{fermi4fgl}), the source is listed with a moderately high variability index of 42. 
    \item[RX~J0812.0+0237] Reported in the 2WHSP catalog with an uncertain redshift of $z\sim 0.2$, a new redshift of $z = 0.1721 \pm 0.0002$ has recently been estimated by \citet{2021MNRAS.504.5258B}. {Its radio and X-ray fluxes are comparable to typical values of 1ES~0229+200.}
    \item[1ES~1028+511] A source with relatively high redshift of $z = 0.360$ \citep{Polomski1997}, chosen on the basis of a high X-ray flux $>10^{-11}$ erg cm$^{-2}$ s$^{-1}$ and a high X-ray-to-GeV-gamma-ray flux ratio of 1.2 among the sample presented in \citet{2020MNRAS.491.2771C}. It was {first} detected at VHE by VERITAS in 2024 \citep{2024ATel16458....1F, 2025arXiv250311543B}, {when also MAGIC and LST-1 observations were carried out.}
    \item[1ES~1426+428] Located at redshift $z = 0.129$ \citep{1989ApJ...345..140R}, it was originally discovered at TeV energies by Whipple \citep{1426_whipple} and the VERITAS telescopes \citep{2017ApJ...835..288A}. A long-term dataset was published by MAGIC in \citet{1st_MAGIC_EHBL_catalog}, while the present work reports additional data from 2020.
    \item[1ES~2037+521] Very nearby source located at $z = 0.053$ \citep{2003A&A...400...95N}, already detected at TeV energies by MAGIC \citep{1st_MAGIC_EHBL_catalog}. A new long-term monitoring of the source was performed between 2018 and 2019, providing valuable insights into its flux variability.

\end{description}
    
\begin{table*}
\centering
\caption{Sample of EHBLs observed with MAGIC. Columns from \textit{left} to \textit{right}: source name, equatorial (RA and DEC, {J2000)} and Galactic coordinates (l and b), redshift $z$, equivalent Galactic hydrogen column density reported by \citet{2005AA...440..775K}, synchrotron peak frequency $\log\nu_{\rm{peak}}$ reported by \citet{3hsp_catalog}, {and the source detection status at VHE gamma rays before this work (Y: detected, N: not detected)}}
\label{tab:source_list}
\renewcommand*{\arraystretch}{1.2}
\begin{tabular}{lccccccccc}
\toprule
\multirow{2}{*}{Source} & RA & DEC & l     & b       & \multirow{2}{*}{$z$} & $N_{H}$& $\log\nu_{\rm{peak}}$ & Known \\
& [$^\circ$] & [$^\circ$]  &  [$^\circ$]     &   [$^\circ$]     & &  $\times10^{20}$\,[cm$^{-2}$]  &   [Hz]  & VHE         \\
\hline
        RBS~42 & 4.62 & 29.79 & 114.45 & -32.54 & $>0.1$\footnote{Uncertain redshift} & 4.06 & 17.1 & N  \\ \hline
        TXS~0637--128 & 100.03 & -12.89 &  223.21 & -8.31 & 0.137 & 31.0 & 17.4 & N \\ \hline
        RX~J0805.4+7534 & 121.36 & 75.57 & 138.9 & 30.8 & 0.121 & 3.28 & 16.3 & N \\ \hline
        RX~J0812.0+0237 & 123.00 & 2.63 & 219.98 & 19.09 & 0.172 & 3.70 & 16.7 & N \\ \hline
        1ES~1028+511 & 157.83 & 50.89 & 161.44 & 54.44 & 0.36 & 1.12 & 16.9 & Y\footnote{\citealt{2024ATel16458....1F}} \\ \hline
        1ES~1426+428 & 217.14 & 42.70 & 77.48 & 64.90 & 0.129 & 0.95  &  18.1  & Y\footnote{\citealt{1426_whipple}} \\ \hline
        1ES~2037+521 & 309.85 & 52.33 & 89.69 & 6.55 & 0.053 & 45.4 & 16.9 & Y\footnote{\citealt{1st_MAGIC_EHBL_catalog}}  \\ \hline
\toprule
\end{tabular}
\end{table*}

\begin{table*}
\centering
\caption{\label{tab:obs_table_magic} Results of the signal search and integral flux analysis of the MAGIC data for the seven EHBLs considered in this study. Columns from \textit{left} to \textit{right}: source name, year(s) of observation, total observation time and effective exposure after quality cuts, signal significance (source highlighted in bold text if detected), assumed energy threshold $E_{\mathrm{th}}$ for the analysis, and the flux measured above $E_{\mathrm{th}}$. In case of non-detection, an upper limit at 95\% confidence level  of the integral-flux was computed (see Section~\ref{sec:MAGIC_results} for details). The source 1ES~1028+511 has been observed with MAGIC-only and also simultaneously with MAGIC and LST-1 (more details in Section~\ref{sec:MAGIC_results}.)}
\begin{tabular}{lccccccc}
\toprule
\multirow{2}{*}{Source}    & \multirow{2}{*}{Observation Period}             & Obs. Time & Eff. Time  & Significance & $E_{\mathrm{th}}$ & Flux$_{\geq E_{\mathrm{th}}}$  \\
&                      & [h] & [h] & [$\sigma$]      & [GeV]       & $\times10^{-12}$ [erg cm$^{-2}$ s$^{-1}$]  \\
\hline
        RBS~42 & 2019 & 43.3 & 20.3 & 0.7 & 200 & $<2.1$  \\ \hline
        TXS~0637--128 & 2017-2025 & 39.9 & 29.1 & $4.3$ & 250 & 1.29 $\pm$ 0.60 \\ \hline
        \textbf{RX~J0805.4+7534} & 2018-2024 & 61.7 & 50.6 & 6.0 & 150  & 4.41$\pm$ 1.40   \\ \hline
        \textbf{RX~J0812.0+0237} & 2019-2020 & 56.8 & 49.0 & 5.6 & 150 & 4.11$\pm$ 0.88 \\ \hline
        \multirow{2}{*}{\textbf{1ES~1028+511}} & [MAGIC-only] 2022-2024 & 43.5 & 25.3 & \multirow{2}{*}{5.7\footnote{Results from the analysis of simultaneous MAGIC+LST1 data.}} & \multirow{2}{*}{150} &  \multirow{2}{*}{4.55 $\pm$ 1.15} \\
         & [MAGIC+LST1] 2022-24 & 15.6 & 6.0 & &  &  \\ \hline
        \textbf{1ES~1426+428} & 2020 & 15.0 & 12.8 & 4.2 & 200 & 7.11 $\pm$ 1.58    \\ \hline \textbf{ 1ES~2037+521} & 2018-2019 & 77.4 & 74.7 & 4.0 & 300 &  0.83 $\pm$ 0.23  \\ \hline
\hline
Total & 2017-2025 & 337.6 & 261.8 \\ 
\toprule
\end{tabular}
\end{table*}


\section{MAGIC Results}
\label{sec:MAGIC_results}
\noindent
The seven targets were observed with the MAGIC telescopes between 2017 and 2025, collecting a total of 338\,h of exposure time. After applying data-quality selections, 262\,h were adopted for the analysis. In particular, specific cuts on the brightest allowed moonlight conditions were applied to ensure a lower energy threshold in the final dataset. Table~\ref{tab:obs_table_magic} summarizes the general information of MAGIC observations. In the following sections, we briefly discuss the results of the analysis of the MAGIC data, whose procedure is reported in detail in Appendix~\ref{app:magic}.


\subsection{Signal Search and Flux Analysis}
\label{subsec:MAGIC_signal}
\noindent
For the signal search, the $\theta^2$ method was adopted. The significance of the gamma-ray signal, calculated with formula [17] of \citet{LiMa83}, is reported in Table~\ref{tab:obs_table_magic}. 

Our analyses reported three new firm (above 5$\sigma$) detections at TeV gamma-ray energies, three targets provided a hint of signal (with significance between 3 and 5$\sigma$), and for one source no significant gamma-ray signal was found in the current dataset.

The two new TeV detections RX~J0805.4+7534 and RX~J0812.0+0237 were both obtained within roughly 50~h of good-quality observations, reporting a similar signal significance and flux intensity compatible with $4\times 10^{-12}$ erg cm$^{-2}$ s$^{-1}$ {above 150~GeV}. The third TeV detection of 1ES~1028+511 is discussed in detail later in this Section. 

The two TeV-detected sources 1ES~2037+521 and 1ES~1426+428 are long-term monitored targets of the MAGIC Collaboration. 
New observations reported in this catalog provide just a hint of detection for both of them, leading to a discussion on their possible long-term flux variability.

In the previous observations of 1ES~2037+521 in 2016 \citep{1st_MAGIC_EHBL_catalog}, a VHE signal was detected with a significance of $7.5\:\sigma$ in 28~h of good-quality data.  In contrast, the more recent dataset from 2018–2019, despite providing a significantly longer good-quality exposure of 77~h, resulted in a lower signal significance of $4.0\:\sigma$.
This difference may be attributed to potential long-term  variability of the source. In particular, during the first observations of 2016, the source was likely to undergo enhanced flaring activity, which would support high signal detection.  
To account for this possible variability, we investigated the presence of day-scale flux variations in the new VHE dataset. This analysis confirms that a constant-flux hypothesis cannot be excluded in the new 2018-2019 dataset ($\chi^2$/d.o.f = 31/31; d.o.f. = degrees of freedom). However, when the investigation is extended to the entire stacked dataset from 2016 to 2019, the observed flux {suggests a moderate tension with the constant-flux hypothesis ($\chi^2$/d.o.f = 58/42, $p_{\text{value}} = 0.05$, with a normalized $\chi^2 \sim 1.35$)}. This long-term variability will be further {discussed along with} the spectral analysis reported in Section~\ref{subsec:MAGIC_spectra}.

The source 1ES~1426+428 is a well-established TeV EHBL which has been monitored by MAGIC in 2010, 2012, 2013 {\citep[see details in][]{1st_MAGIC_EHBL_catalog}}, and 2020 {(this work)}, showing possible hints of variability of the TeV flux. In fact, a {clear} detection of the source was only {found} in the 2012 dataset, while in the other datasets - including this new dataset from 2020 - just a hint of signal {has been identified}. We find that the constant-flux hypothesis {is still acceptable} over the 2020-only dataset ($\chi^2$/d.o.f = 13/10, $p_{\text{value}} = 0.22$), and also for the stacked 2010-13 + 2020 dataset  ($\chi^2$/d.o.f = 15/18). Interestingly, for the 2020-only dataset - which does not report a {clear detection at VHE above 5$\sigma$} - the flux analysis yields a rather high value of $(7.11 \pm 1.58) \times 10^{-12}$ erg cm$^{-2}$ s$^{-1}$. This value is slightly higher than the average value found in the first MAGIC catalog of extreme blazars \citep{1st_MAGIC_EHBL_catalog}, but it is compatible with the higher flux reported for the 2012-only dataset, suggesting a new possible high state of the source at these energies.

The third hint of detection is reported here for TXS~0637--128. This source was already observed between 2016 and 2017 in the first MAGIC catalog of extreme blazars \citep{1st_MAGIC_EHBL_catalog}, eventually leading to a low signal significance of 1.7$\sigma$ after 12.8~h of observations, \lf{mainly performed under moderate Moonlight conditions}. In this work, new observations were performed between 2018 and 2025 under dark conditions and with a lower energy threshold, resulting in  additional 29.1~h of good-quality data.. The analysis of this new dataset results in a stronger signal of 4.3$\sigma$ significance with respect to the previous dataset, although still not sufficient to claim a firm detection of the source. The flux analysis reports compatibility with a steady emission state ($\chi^2$/d.o.f = 21/25). We also investigated whether stacking the earlier 2016–2017 dataset with the new 2018–2025 observations would enhance the signal significance.  However, the combined analysis of the two datasets yields a comparable signal significance of 3.3$\sigma$ over 45~h of good-quality data (selected by applying uniform quality cuts), likely due to an increase in background events.

Additionally, we report no significant signal from one TeV candidate, named RBS~42.
It is important to mention that the signal detection at TeV energies is strongly affected by the energy threshold resulting from the observational configuration. Indeed, this non-detected source was observed under Moonlight \citep[more details on ][]{magicperf_moon:2017}, leading to a high energy threshold which may have reduced the gamma-ray signal detectability. \\

\subsubsection{MAGIC and LST-1 joint observations of 1ES~1028+511}
\noindent
In 2024 the source 1ES~1028+511 was observed by the VERITAS Collaboration {for 49~h}, which announced its detection at TeV gamma rays in \citet{2024ATel16458....1F} and later in \citet{2025arXiv250311543B}. 
The observed photon index is $\Gamma = 3.6 \pm 0.5$, and the photon flux above 200~GeV  is $F(>\text{200 GeV}) = (2.4 \pm 0.5_{\text{stat}}) \times 10^{-12}$  cm$^{-2}$ s$^{-1}$.

The MAGIC Collaboration also observed the source over the period from 2022 to 2024, for a total of 25.3~h after quality cuts. During this period, the analysis of  MAGIC-only data reports no significant signal detection above 150 GeV. Applying the same energy threshold of 200 GeV adopted in the analysis by VERITAS, we provide an integral upper limit at 95\% confidence level of $6.4 \times 10^{-12}$ cm$^{-2}$ s$^{-1}$, which is consistent with their integral photon flux reported above. 

\begin{figure}
\centering
\caption{Event $\theta^2$ distribution from the direction of 1ES~1028+511 resulting from the overall dataset obtained with the MAGIC and LST-1 telescopes {between 80 and 500 GeV}. 
The gamma-ray like events are represented by the blue markers, while the background is denoted by orange markers. The vertical dashed line indicates the defined signal region $\theta^2 < 0.04$ deg$^2$ from which the significance of the detection is calculated. }
\label{Fig:theta2_joint_1ES1028}
\includegraphics[width=0.95\columnwidth]{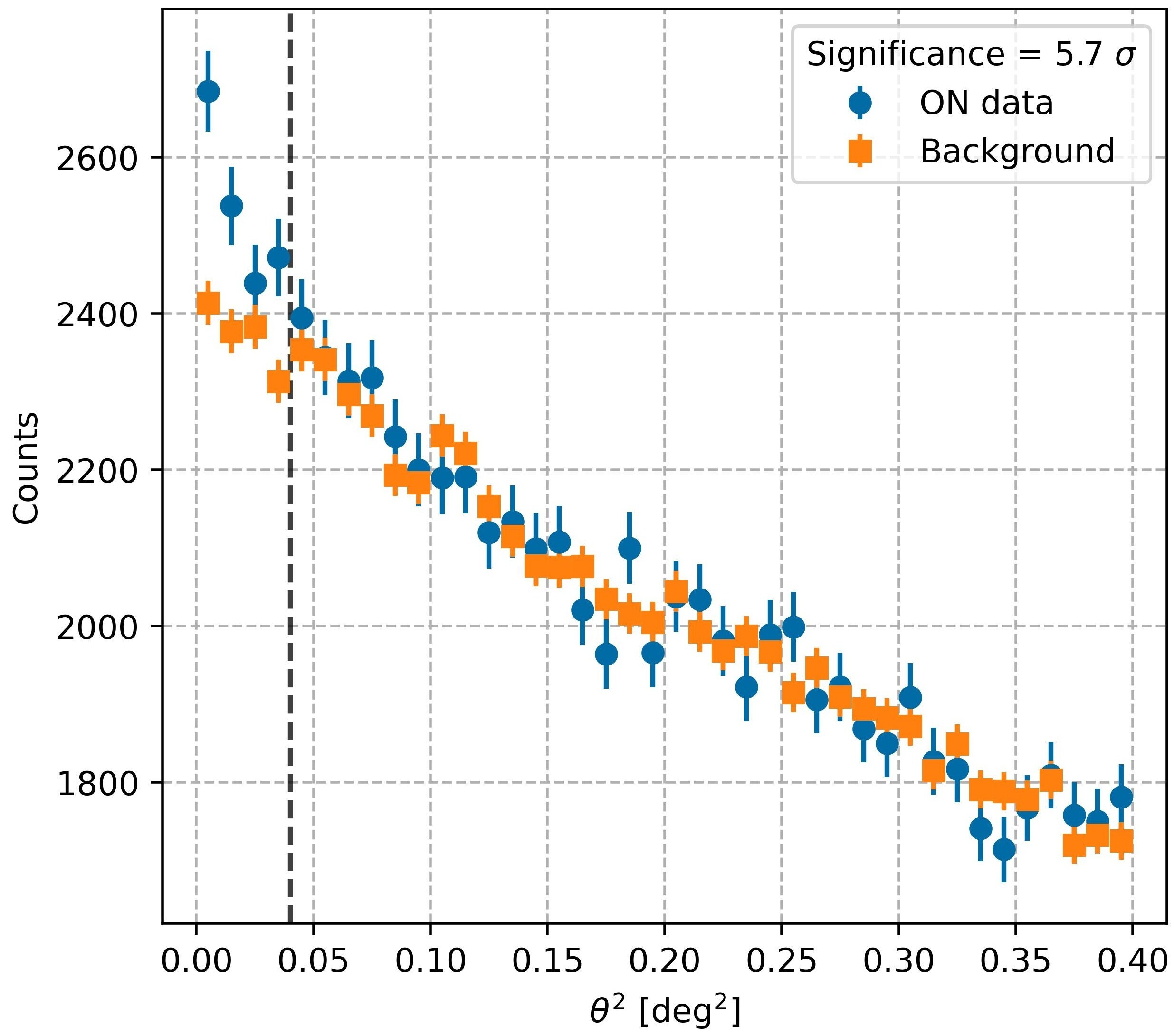}
\end{figure}

During the MAGIC observations of 1ES~1028+511, {joint LST-1 observations were also performed for a total of $15.6$~h {(see \Cref{tab:obs_table})}. This simultaneous MAGIC and LST-1 data restricts to 6.0~h  after quality cuts spread over 7 nights between 2022 and 2024.}
Thanks to the recent development of a joint analysis procedure, we performed the analysis of this simultaneous dataset (6.0~h), which was joined to the remaining data from MAGIC-only non-coincident observations of (19.3~h out of the 25.3~h previously mentioned). Details of the joint MAGIC and LST-1 dataset are reported in Appendix~\ref{sec:LST_data_analysis_details}. 
The signal search analysis of this overall joint dataset reports a signal of 5.9~$\sigma$ between 80 and 500 GeV, which confirms the previous TeV detection of the source reported by VERITAS. The corresponding $\theta^2$ plot is shown in \Cref{Fig:theta2_joint_1ES1028}.

The detection of the source obtained with these joint MAGIC and LST-1 observations  highlights the improved sensitivity of the new generation of Cherenkov telescopes, represented in this case by the LST-1. 
However, a comparison between MAGIC-only and MAGIC+LST-1 results would be affected by differing systematic uncertainties between the data analysis pipelines, and thus cannot be made directly without properly accounting for those effects.

The integral photon flux  of this source above the energy threshold of 150 GeV resulting from our overall dataset is $(4.5 \pm 1.1) \times 10^{-12}$ cm$^{-2}$ s$^{-1}$, compatible with the previously computed upper limit.  
In order to compare with VERITAS' results, we also computed the integral photon flux above 200 GeV, which is $(2.2 \pm 0.9) \times 10^{-12}$ cm$^{-2}$ s$^{-1}$, in agreement with VERITAS findings. \\

\setlength{\tabcolsep}{3pt}
\begin{table}
\centering
\caption{Results of the MAGIC spectral analysis of the EHBLs detected at VHE gamma rays {with more than 4$\sigma$ statistical significance (see \Cref{tab:obs_table_magic}).}  For each source, we report $E_{\mathrm{dec}}$, differential {photon} flux $F_0$ derived from the observed spectrum at the decorrelation Energy $E_{\mathrm{dec}}$, spectral indices of the observed  $\Gamma_{\mathrm{obs}}$ and of the intrinsic EBL-deabsorbed spectrum $\Gamma_{\mathrm{int}}$. Only statistical errors are reported.}
\renewcommand*{\arraystretch}{1.2}
\hspace*{-45pt}\begin{tabular}{lcccc} 
\toprule
\multirow{2}{*}{Source}      & $E_{\mathrm{dec}}$ & $F_0$& \multirow{2}{*}{$\Gamma_{\text{obs}}$} & \multirow{2}{*}{$\Gamma_{\text{int}}$}\\
               &[TeV]  &[TeV$^{-1}$ cm$^{-2}$ s$^{-1}$] & & \\
\hline
RX~J0805.4+7534    &  0.93 & $(3.9\pm 1.2) \times 10^{-13}$ & $2.8  \pm 0.4$ & $2.0  \pm 0.4$\\
RX~J0812.0+0237   &  0.32 & $(5.2\pm 1.0) \times 10^{-12}$ & $2.9\pm 0.3$ & $2.5\pm 0.3$\\ 
1ES~1028+511\footnote[2]{Results from the analysis of joint MAGIC and LST-1 data.}  &  0.14 & $(1.5 \pm 0.3) \times 10^{-10}$ & $3.5\pm 0.5$ &  $2.3\pm 0.5$\\
1ES~1426+428$^a$  &  0.72 & $(2.1\pm 0.4) \times 10^{-12}$ & $2.5\pm 0.3$ & $1.7\pm 0.3 $   \\ 
1ES~2037+521\footnote[1]{Only hint of signal was detected for these sources, see \Cref{tab:obs_table_magic}} &  0.48 & $(1.3\pm 0.4) \times 10^{-12}$ & $2.4\pm 0.3$ & $2.1\pm 0.3$\\
\toprule
\end{tabular}
\label{tab:spectra_magic} 
\end{table}

\subsection{Spectral Analysis}
\label{subsec:MAGIC_spectra}
\begin{figure}
\centering
\caption{{SEDs} of RX~J0805.4+7534, RX~J0812.0+0237, 1ES~1028+511,  1ES~1426+428, and 1ES~2037+521. Empty circles indicate observed data, filled circles refer to EBL-corrected data. Comparative datasets from previous observational campaigns are reported in light gray markers.}
\label{fig:MAGIC_spectra_all}
\includegraphics[width=0.95\columnwidth]{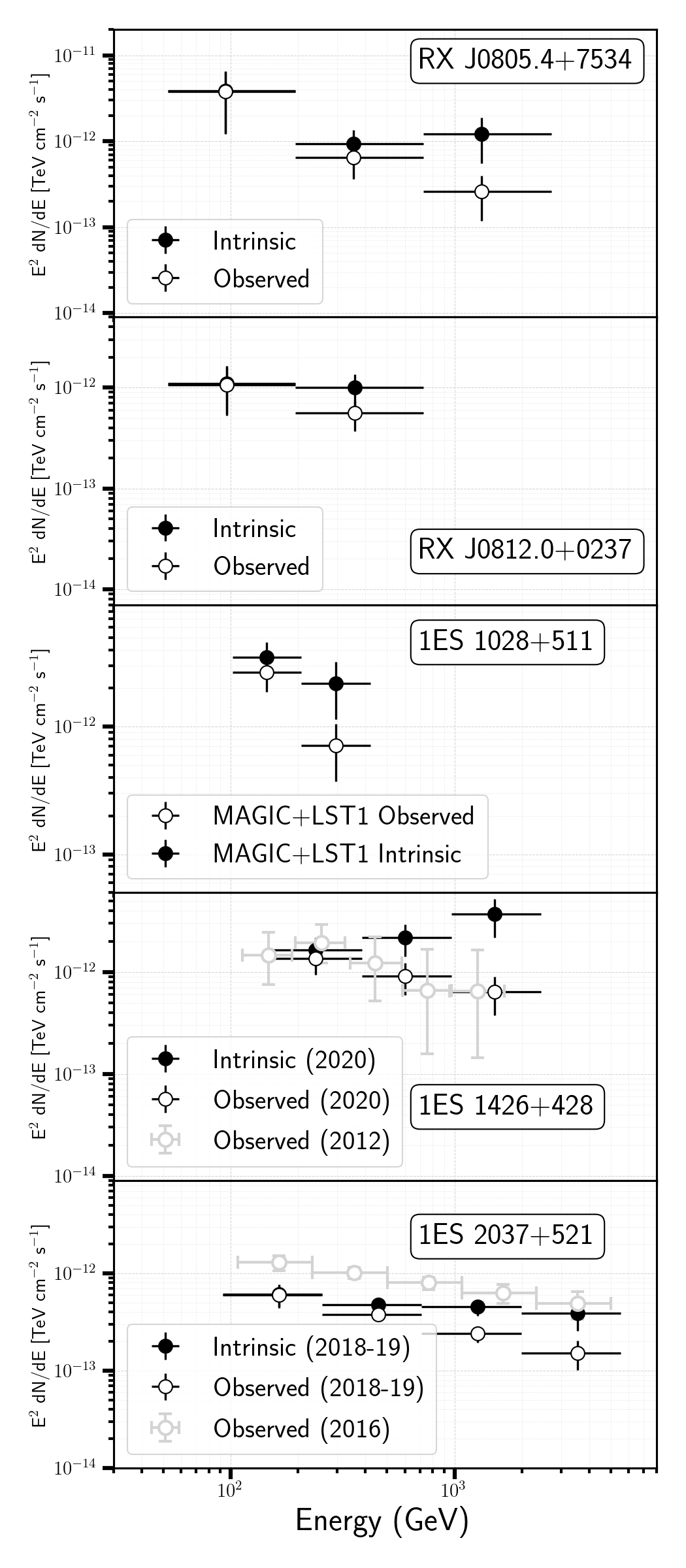}
\end{figure}

\noindent
For five sources, we performed a spectral analysis, reporting the results in \Cref{tab:spectra_magic} and {in \Cref{fig:MAGIC_spectra_all}}. Among them, we have included those with a new MAGIC detection or with a hint of detection at VHE gamma rays for already known TeV emitters. 
The spectra are fit with a simple power law described by:
\[
    \frac{dN}{dE} = F_0 \: \left( 
    \frac{E}{E_{\rm{dec}}}\right)^{-\Gamma},
\]
where $F_0$ is the flux normalization and  $E_{\rm{dec}}$ is the decorrelation energy (energy corresponding to the minimum correlation between flux normalization and spectral index, computed according to \citealt{2010ApJ...708.1310A}). 

All observed spectra, characterized by the photon index $\Gamma_{\text{obs}}$, were also corrected for EBL attenuation to estimate the intrinsic emission of each source, described by $\Gamma_{\rm{int}}$. Throughout the paper, the EBL was accounted for with the model by \citet{Dominguez:2010bv}.

For TXS~0637--128 and RBS~42, which are sources without VHE detection, we calculated flux upper limits at 95\% confidence level assuming a typical observed photon index of 2 (see Table\,\ref{tab:obs_table_magic}). We investigated softer photon indices up to 3, which would lead to different flux upper limits, even though always within the instrument systematic uncertainties ($<15\,\%$).

For 1ES~2037+521 and 1ES~1426+428, {whose detection at VHE was already reported} by the MAGIC Collaboration in \citet{1st_MAGIC_EHBL_catalog} but that are not clearly detected in this new dataset, we have extracted the spectral information in order to compare it with the previous observation periods and investigate the possible long-term spectral variability of the sources. 
Interestingly, the comparison of the new datasets -- also reported in \Cref{fig:MAGIC_spectra_all} -- shows different results with respect to past observations. In the case of 1ES~2037+521, we find a spectrum with compatible photon index but different average flux, probably due to variability of the source.  On the other hand, the new data for 1ES~1426+428 in 2020 report a perfect compatibility with the spectrum observed in 2012, which was also showing compatible integral flux, as discussed above.

For the source 1ES~1028+511, we extracted the photon index from the analysis of joint MAGIC and LST-1 data, as described in Appendix~\ref{sec:LST_data_analysis_details}. The observed photon index results in $\Gamma = 3.5 \pm 0.5$ and the intrinsic one after EBL de-absorption is $\Gamma = 2.3 \pm 0.5$, which is in agreement with VERITAS findings. 

Finally, it is worth to notice that among all sources investigated here, most of them have shown a soft intrinsic  TeV  spectrum. Interestingly, only the source 1ES~1426+428 has shown a clearly hard-TeV intrinsic spectrum, a result that is in agreement with previous findings for this source. 
These results align with the different nature of extreme blazars at VHE discussed above, where EHBLs can show similar broadband properties but eventually are characterized as soft- and hard-TeV spectra.

\section{{\it Fermi}-LAT results}
\label{sec:Fermi_results}
\noindent
The HE gamma-ray range is essential in constraining the interpretation of the spectral properties of EHBLs at VHE energies. In this work, this information was provided by the \textit{Fermi}-LAT data, which were analyzed as described in Appendix~\ref{app:fermi_analysis}. 

A first attempt of analyzing only data strictly simultaneous to the MAGIC observations was performed. However, most of the sources were not  detected within these individual short-time datasets.
Then, in order to provide comparable results, an analysis covering all data available at the time of the analysis was conducted, integrating all {\it Fermi}-LAT exposure on the source from the beginning of the operations in August 8$^{\text{th}}$, 2008 up to December 25$^{\text{th}}$, 2024.

\Cref{tab:spectraLATandXRT} reports the results of our analysis. All our targets are detected over the background photons (test statistics TS~$> 25$, where the square root of the TS is approximately equal to the detection significance $\sigma$ for a given source, \citealt{TSpaper}), some of them with a very strong signal (e.g. RX~J0805.4+7534). 
These results, as mentioned before, are in agreement with the {phenomenology} of extreme blazars. Depending on the prominence of the high-energy hump in the HE gamma-ray regime, different detectability for the {\it Fermi}-LAT instrument is naturally expected. 

Our analysis results are evaluated in relation to the most recent catalogs from the \textit{Fermi}-LAT Collaboration, namely the 4FGL catalog \citep{fermi4fgl} and 3FHL catalog \citep{fermi3FHL}. \Cref{tab:fermi_lat_catalog_info} highlights the most relevant information to be compared.
The variability index of the 4FGL catalog identifies sources likely to be steady when lower than 27.69. In our source list, most of them are classified as non-variable objects. An exception is found for RX~J0805.4+7534, 
with a value quite above this limit as the source is possibly showing long-term variability in this energy band.
Our resulting photon indices listed in \Cref{tab:spectraLATandXRT} are compatible with the values reported in the 4FGL catalog. On the other hand, the photon flux is not directly comparable, as for the \textit{Fermi}-LAT catalogs the integration is within 10-100~GeV and in our case we integrate from 10 to 150~GeV. However, an evident difference can be identified for the photon flux of RX~J0805.4+7534, possibly due to its variability with respect to the other objects or different spectral characteristics (see \Cref{tab:fermi_lat_catalog_info}). For the three long-term monitored sources TXS~0637--128, 1ES~2037+521, and 1ES~1426+428, our results are also compatible with our previous analyses reported in \citet{1st_MAGIC_EHBL_catalog}, even if there  specific integration times for each source were adopted.

\begin{table*}
	\centering
		\caption{\label{tab:spectraLATandXRT}Spectral parameters resulting from the analysis of {\it Swift}-XRT and {\it Fermi}-LAT data. Columns from \textit{left} to \textit{right}: source name; {\it Swift}-XRT data analysis time interval, 2-10 keV X-ray flux; spectral index; {\it Fermi}-LAT data analysis time interval; 1-150 GeV HE gamma-ray photon flux; spectral index; likelihood test statistics (TS) of the fitted model. The square root of the TS is approximately equal to the detection significance $\sigma$ for a given source.}	
\setlength{\tabcolsep}{0.36em}
\begin{tabular}{l|ccc|cccc} 
    \hline
   &\multicolumn{3}{c|}{{\it Swift}-XRT} & \multicolumn{4}{c}{{\it Fermi}-LAT}  \\ 
    \cline{2-8}
    Source & Obs. date & F$_{(2-10\,\text{keV})}\times$10$^{-12}$  & \multirow{2}{*}{$\Gamma$} & Interval & F$_{(1-150\,\text{GeV})}\times$10$^{-10}$ & \multirow{2}{*}{$\Gamma$} & \multirow{2}{*}{TS}  \\
            & [MJD] & [erg cm$^{-2}$ s$^{-1}$]& & [MJD]   & [ph cm$^{-2}$ s$^{-1}$]   & & \\
    \hline
    RBS~42	&	54245-58533	&	6.5$\pm$2.5	&	2.02$\pm$0.06	&	54682	-	60670&	11.2	$\pm$	1.3	&	1.66	$\pm$	0.08	&	223	\\ \hline
    TXS~0637--128	&	55052-59806	&	35.8 $\pm$6.9 &	1.77$\pm$0.10	&		54682	-	60670			&	19.7	$\pm$	2.1	&	1.69	$\pm$	0.08	&	259	\\ \hline
    RX~J0805.4+7534	&	55232-58584	&	56.2$\pm$35.2	&	2.10$\pm$0.10	&	54682	-	60670&	105.0	$\pm$	3.0	&	1.85	$\pm$	0.02	&	4069	\\ \hline
    RX~J0812.0+0237	&	58485-58880	&	2.8$\pm$1.3	&	1.98$\pm$0.15	&	54682	-	60670 &	26.0	$\pm$	2.4	&	1.79	$\pm$	0.07	&	385	\\ \hline
    1ES~1028+511	&	54614-60375	&	16.8$\pm$6.3	&	2.08$\pm$0.12	&		54682	-	60670	&	47.6	$\pm$	2.0	&	1.69	$\pm$	0.08	&	2134	\\ \hline
    1ES~1426+428	&	57781-59132	&	2.5$\pm$1.0 &	1.85$\pm$0.11	&	54682	-	60670&	22.7	$\pm$	1.3	&	1.57	$\pm$	0.08	&	1322	\\ \hline
    1ES~2037+521	&	58218-58753	&	4.6$\pm$2.4	&	1.94$\pm$0.30	&	54682	-	60670&	27.5	$\pm$	2.7	&	1.78	$\pm$	0.08	&	262	\\ \hline
\hline
\end{tabular}
\end{table*}

\section{Swift results}
\label{sec:Swift_results}

\subsection{XRT instrument}
\label{sec:X-ray_prop}
\noindent
EHBLs are typically bright sources in the X-ray band. This feature is connected with their classification based on a synchrotron peak above $10^{17}$~Hz $\simeq 0.3$ keV, implying that most of their synchrotron emission is  in the X-ray band. For this reason, X-ray observations are crucial for the study of EHBLs.

In this paper, we have obtained simultaneous observations with the X-ray Telescope \citep[XRT,][]{Burrows04} onboard the {\it Neil Gehrels Swift Observatory} through Target of Opportunity (ToO) requests. Additionally, all available {\it Swift}-XRT archival data \citep{2013ApJS..207...28S} were analyzed following the procedure outlined in Appendix~\ref{app:xrt_analysis}.

\begin{figure*}
\centering
\caption{\label{Fig:Xray_LC}\textit{Left panels:} X-ray light curves (2-10\,keV), corrected for Galactic extinction. Shadowed areas illustrate MAGIC observation windows. \textit{Right panels:} Scatter plots of the power-law photon index $\Gamma$ versus X-ray flux (2-10\,keV) measured with {\it Swift}-XRT for each source of the sample. Dashed red lines represent the best-fit linear models for each source. When the X-ray observations are more than five, we also indicate the resulting Pearson coefficient to quantify the correlation between the points, indicating possible \textit{harder-when-brighter} behaviors. In the case of 1ES~1426+428, we have split the dataset in two periods MJD 54500--55500 (blue points) and MJD 57500--59500 (lightblue points), as discussed in the text.   }
\includegraphics[width=0.89\textwidth]{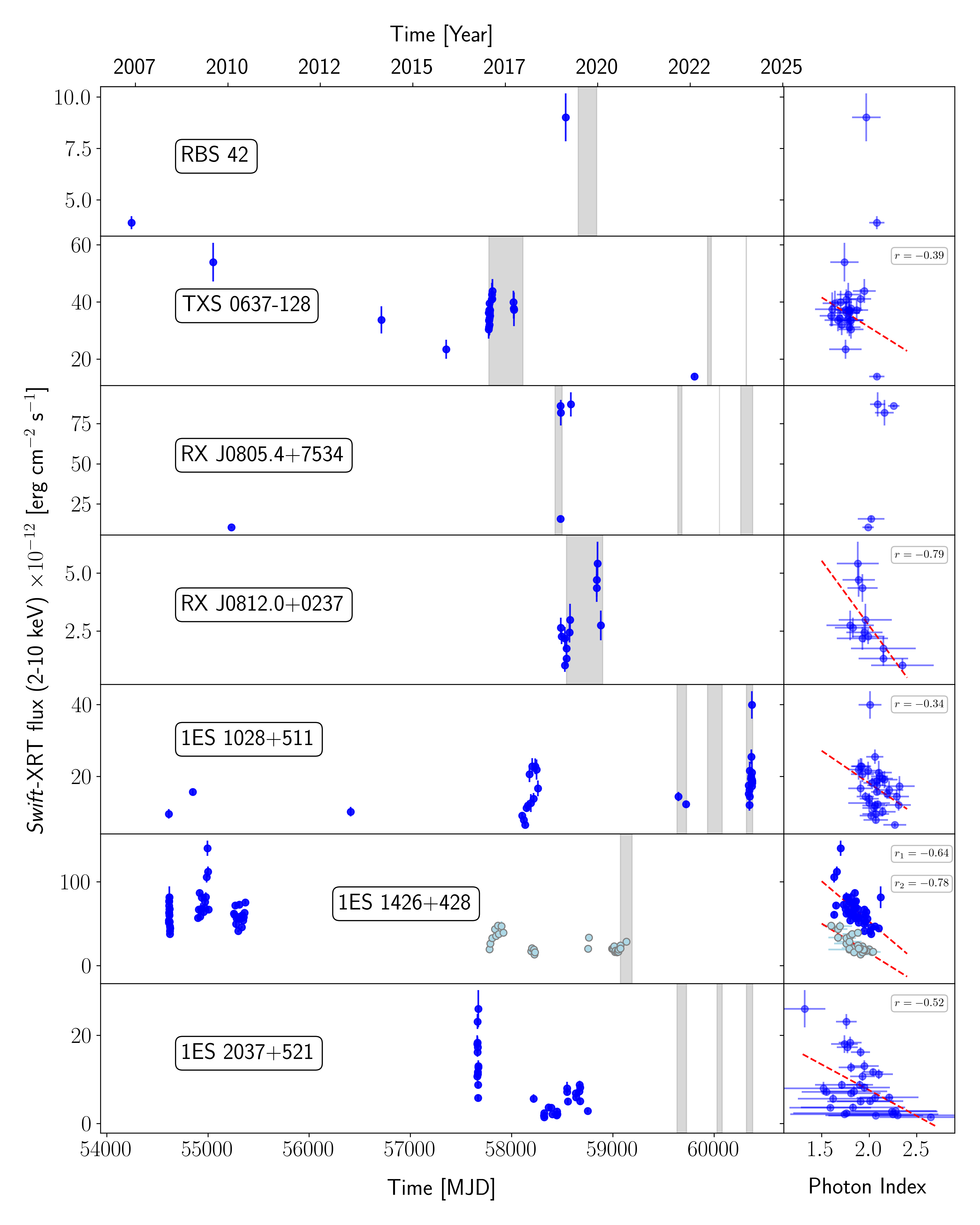}
\end{figure*}

The results are summarized in \Cref{tab:spectraLATandXRT} and the light curves in the 2-10 keV energy range are shown at the left of Figure~\ref{Fig:Xray_LC}. 
A power-law spectrum was adopted for the spectral analysis. In this work, we merge the new \textit{Swift} datasets with those reported in the first MAGIC EHBL catalog \citep{1st_MAGIC_EHBL_catalog}, in order to explore also the long-term variability of these sources.

Most of the sources of this catalog exhibit a photon index compatible ({in the case of} 1ES~1028+511, RX~J0805.4+7534, and RBS~42) or harder (RX~J0812.0+0237, TXS~0637--128, 1ES~2037+521 and 1ES 1426+428) than $\Gamma \sim 2$ in this energy band {(very hard for TXS~0637--128)}, indicating that the synchrotron peak was located in or above this energy range during these ToO observations.

The relation between integral flux and spectral index in the X-ray range for these sources was investigated. As shown in the right panel of \Cref{Fig:Xray_LC}, a quite common \textit{harder-when-brighter} behavior emerges, indicating an anti-correlation between a decreasing photon index and an increasing flux. 
Where the X-ray observations provide sufficient statistics (more than five data points), we also indicate the resulting Pearson coefficient $r$ to quantify the correlation between the points. It is worth noticing that the correlation analysis does not incorporate error bars, and thus the results represent an indication rather than a firm assessment.

The \textit{harder-when-brighter} trend is statistically supported with a high correlation ($|r| > 0.7$) in the dataset of RX~J0812.0+0237. 
A moderately high correlation is also found in the case of 1ES~1426+428. Across the full dataset, it is the most variable source in our sample, reaching flux variations of about two orders of magnitude. The flux-versus-photon-index plot suggests a possible clustering of the data in two different periods of activity, which we identify as MJD 54500--55500 and MJD 57500--59500. Accordingly, we have performed two calculations of the correlation index, resulting in $r_1 = -0.64$ and  $r_2 = -0.78$, respectively, which confirm a moderately high harder-when-brighter trend in both activity states.
The source 1ES~2037+521 shows an increased X-ray flux variability between the observations before and after 2017. Nevertheless, the correlation index of $|r| \sim 0.5$ could partly arise from a phase of spectral softening combined with a lower detection significance in this energy band.

These results are typical in blazars and describe an increasing synchrotron-peak energy when the flux increases. An explanation for this phenomenon is an increasing maximum particle energy distribution during high-state activities, observed also for Mrk~501 in several X-ray campaigns \citep{1998ApJ...492L..17P}. However, this is not always verified in flaring blazars, as indicated by observations of Mrk~501 in 2012, when the source exhibited very hard spectra in the X-ray and VHE ranges both in a quiescent and a flaring state \citep{2018AA...620A.181A}.

In the current selected X-ray datasets, the overall observations are never compatible with the constant-flux hypothesis. This confirms that the long-term activity of the AGNs steadily affects the high-energy tail of the particle distribution lying in this energy band.
Even taking into consideration the datasets quasi simultaneous to MAGIC observations reported with shadowed areas in \Cref{Fig:Xray_LC}, we notice that most of the sources are showing moderate to high intrinsic variability (particularly high for example in the case of RX~J0805.4+7534), which also affects the modeling reported later in the SEDs shown in \Cref{fig:mwl_sed}.

\subsection{UVOT instrument}
\label{sec:UV_prop}
\noindent
For all sources, an analysis of \textit{Swift}-UVOT data was performed, as detailed in Appendix~\ref{app:uvot_analysis}.
The resulting optical-UV light curves of the \textit{Swift}-UVOT instrument are shown in \Cref{Fig:UVOT_LC}. 

For blazars in general, the non- thermal continuum in this energy band usually dominates over the thermal emission of the host galaxy.
Conversely, in EHBLs the synchrotron emission is pushed to higher energies, making it easier to identify the host galaxy emission, and consequently, facilitating the redshift determination through spectroscopic measurements.

Differently from the X-ray band, these optical-UV light curves show just a moderate variability, which is expected due to the longer cooling timescales for the synchrotron emission in this energy range.\\

\section{Tuorla optical results}
\label{sec:tuorla_results}
\begin{figure*}
\centering
\caption{Optical light curves, corrected for Galactic extinction, obtained with \textit{Swift}-UVOT,  KVA (until the end of 2019), and TJO (starting from 2021). Shadowed vertical bands illustrate MAGIC observation windows. The specific filters available for each source are reported in the legend.}
\includegraphics[width=0.9\textwidth]{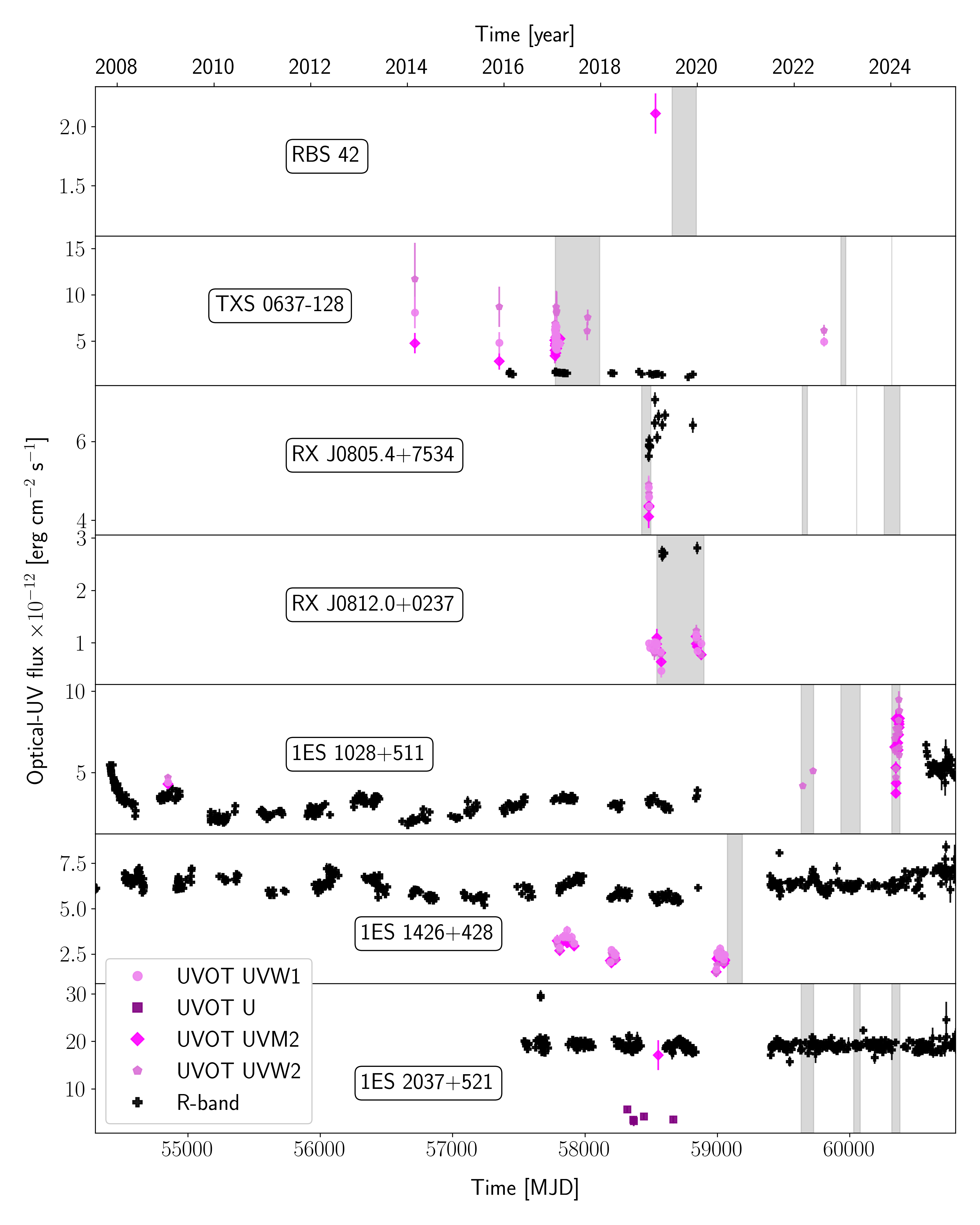}
\label{Fig:UVOT_LC}
\end{figure*}

\noindent
Some bright sources of this sample are also part of the \textit{Tuorla blazar monitoring program} in the optical R-band \footnote{https://tuorlablazar.utu.fi/}. Observations were performed using the Kungliga Vetenskaps Akademi (KVA) telescope until the end of 2019 and with Joan Oró Telescope (TJO) starting from 2021. Data were analyzed using the semiautomatic pipeline \citep{2018A&A...620A.185N}. In \Cref{Fig:UVOT_LC}, we also show the observed fluxes obtained by this optical monitoring, already  corrected for galactic extinction using the same procedure as for the UVOT data.

Similarly to UVOT light curves, just a moderate variability is generally identified. A possibly flaring activity is shown by the data in the case of 1ES~1028+511 in 2008 and also in the next data of 2025, even though no simultaneous VHE observations are available.\\

\section{Broadband analysis and modeling}
\label{sec:OtherBands}
\noindent
The MWL SEDs of all seven targets are shown in \Cref{fig:mwl_sed}. For each source, we show the simultaneous \textit{Swift} data and the new analysis of \textit{Fermi}-LAT data presented in the previous sections. We also include the archival data available at the ASI Space Science Data Center (SSDC)\footnote{\href{http://www.asdc.asi.it}{http://www.asdc.asi.it}}, gathering data at radio frequencies, optical-UV, X-rays, and the most updated HE gamma-ray data of the 4FGL \textit{Fermi}-LAT catalog. 


\label{sec:modelling}
In \Cref{fig:mwl_sed}, we also provide a modeling for the sources newly detected and those with a hint of detection at VHE gamma rays in this dataset, as reported in \Cref{tab:spectra_magic}. Furthermore, we provide a tentative modeling of TXS~0637--128, since the overall dataset quite constrain the parameter space of the model. Conversely, concerning the source RBS~42, although long-term \textit{Fermi}-LAT data and VHE gamma-ray upper limits are available, the lack of MWL data prevents a proper constraint of the synchrotron component. Therefore, we do not provide any modeling for this source, as the parameter space would remain largely unconstrained.

The modeled data of \textit{Swift}-XRT and UVOT were selected when mostly simultaneous with MAGIC observations (shortest gap between the observations), which are marked with shaded gray bands in \Cref{Fig:Xray_LC}. Unfortunately, no simultaneous \textit{Swift} and MAGIC observations are available for 1ES~2037+521. However, as a reference point, we adopted the flux of the earlier X-ray observations in order to apply the theoretical model. 
For 1ES~1426+428, the flux points obtained during simultaneous MAGIC observations are compatible, and consequently, we produce just a SED of XRT data obtained on MJD~55360, as a representative for the whole period.
Conversely, RX~J0812.0+0237, 1ES~1028+511 and RX~J0805.4+7534 showed high and moderate flux variations during the simultaneous observations with MAGIC, respectively. For this reason, we chose two SEDs for each source to represent the states of minimum and maximum flux activity (MJD~58483 and MJD~58484 for the former, MJD~58542 and MJD~58850 for 1ES~1028+511, and MJD~55232 and MJD~58584 for the latter). However, considering that simultaneous MAGIC observations included all these possibly different states, integrating most of the signal from the highest states, we modeled the highest flux emission in X-rays.
In the case of TXS~0637--128, the combined effects of a nearby star and uncertainties in the Galactic extinction estimation make it difficult to obtain a precise determination of the unabsorbed UV flux \citep[see also][]{2020MNRAS.497...94P,2018A&A...620A.185N}. Given the complexity of this optical–UV field, we do not apply a host galaxy model and instead anchor the SED modeling primarily to the X-ray and gamma-ray data.

Given the low flux variability in HE gamma~rays, we adopted the long-term \textit{Fermi}-LAT datasets for the modeling of these sources.  In the case of RX~J0805.4+7534, which shows a modest indication of long-term variability in the 4FGL catalog (see \Cref{tab:fermi_lat_catalog_info}), we also used the long-term dataset, as it provides a valuable counterpart to the MAGIC observations collected over several years and these results are in substantial agreement with those reported in the catalog.

In this work, we apply the standard one-zone leptonic model -- implemented through the \texttt{agnpy} code -- which has been historically used to interpret the MWL emission of BL Lac objects. 
Its application is motivated by the relatively soft TeV spectra of the sources in our sample, which make this scenario particularly suitable, rather than adopting more complicated models with larger parameter space.
In this scenario, relativistic electrons are accelerated within the relativistic jet of the source, moving with a bulk Lorentz factor $\Gamma$ at an angle $\theta$ with respect to the observer's line of sight. They are assumed to fill a spherical region of radius $R$ with density $N$, permeated by a uniform magnetic field $B$. Special relativistic effects are described by the Doppler factor $\delta = [\Gamma(1-\beta \cos \theta)]^{-1}$.

The population of relativistic electrons is assumed to be described by a broken power-law distribution as a function of the electron Lorentz factor $\gamma$:
\[
N(\gamma)=\,K \,\gamma^{-p_1}\, \Big(1+\frac{\gamma}{\gamma_b}\Big)^{p_1-p_2}\; ,
\]
where $K$ is the normalization factor, and $p_1$ and $p_2$ are the indices below and above the break, respectively, with $\gamma_b$ representing the break Lorentz factor.
These electrons emit synchrotron radiation, which is subsequently reprocessed via inverse Compton process, generating the high-energy SSC hump. 

The one-zone SSC model adopted in this work generally successfully reproduces the emission of most sources in our sample, particularly those characterized by a relatively soft TeV spectrum. In the case of 1ES~2037+521, the moderate flux gap between the HE and the VHE gamma-ray data is likely due to long-term variability of the source and the different integration times of the \textit{Fermi}-LAT and MAGIC telescopes data. Conversely, in the case of 1ES~1426+428, this feature is additionally complicated by a very hard intrinsic VHE spectrum, which the model underestimates. These results support the suitability of one-zone SSC model for soft-TeV EHBLs, while highlighting its limitations in modeling hard-TeV EHBLs (like 1ES~1426+428), where more complex scenarios may be required.

The SSC model parameters derived for our sample of EHBLs - reported in \Cref{tab:modeling} - are broadly consistent with typical values of this class of objects. Moderately high bulk Lorentz factors ranging from 10 to 30 indicate significant Doppler boosting of the emission region. A distinctive feature of EHBLs is the very low magnetic field, which leads to inefficient synchrotron cooling, allowing particles to be accelerated to very high energies.
In our sample, the magnetic field lies between 0.06 and 0.2~G, which represent low values but still in agreement with the typical results obtained in the modeling of HBLs. These findings suggest that the sources analyzed here belong mainly to the soft-TeV EHBL class, rather than to the most extreme hard-TeV EHBLs, where magnetic field values can drop to the mG level \citep{Tavecchio2011, Costamante2018}. The relatively high minimum electron energy, which supports the spectral components pushed at very high energies, is also a common feature observed in EHBLs. \\

\begin{figure*}
\centering
\caption{Broadband SEDs for all sources observed by MAGIC and presented in the study. {Optical points from the Tuorla observatory and \textit{Swift}-XRT/UVOT} points shown here are only the datasets quasi simultaneous to MAGIC observations considered for the modeling. Gray markers are archival data from the ASI Space Science Data Center (SSDC) website. Black continuous and dashed lines represent the outcome of the synchrotron and SSC spectra resulting from the application of the single-zone SSC model on the intrinsic EBL-deabsorbed data of the TeV-detected sources. For most of the sources, we also show an estimation of the host galaxy emission {based on the estimated or suggested redshift}, following the model by \citet{hostgalaxymodel}.  
}
\label{fig:mwl_sed}
\subfloat[RBS~42.]{\includegraphics[width=0.5\textwidth]{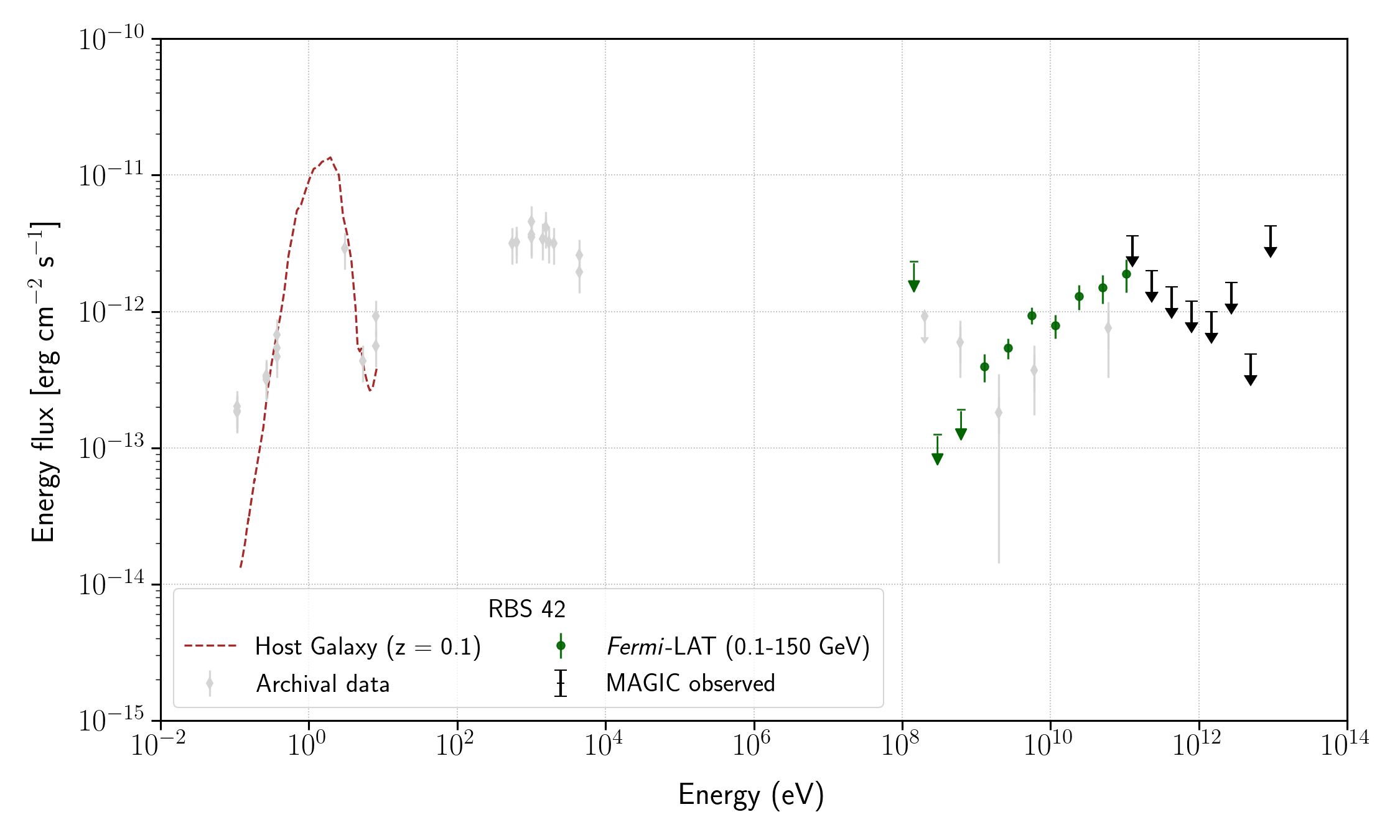}}
\subfloat[TXS~0637--128.]{\includegraphics[width=0.5\textwidth]{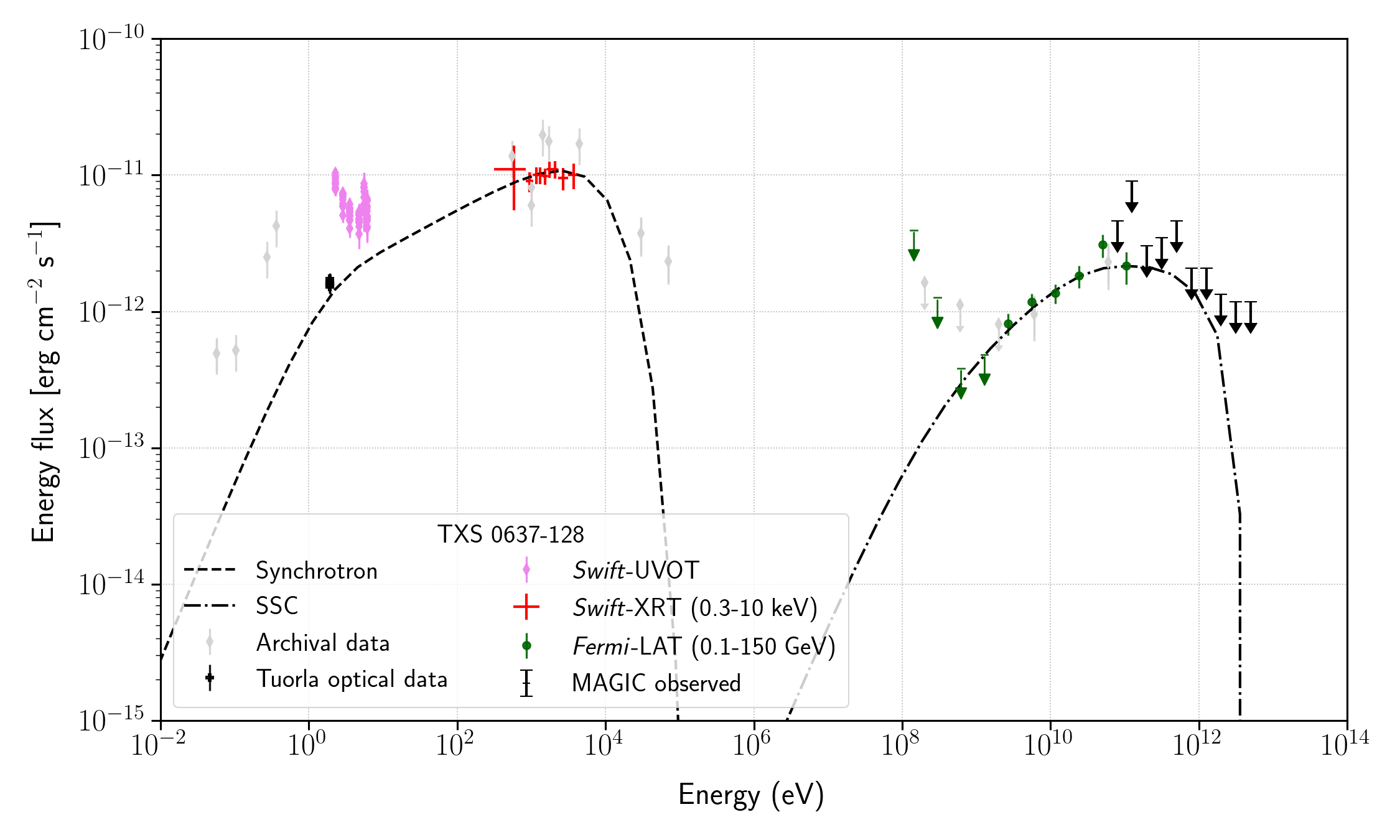}}\\
\subfloat[RX~J0805.4+7534.]{\includegraphics[width=0.5\textwidth]{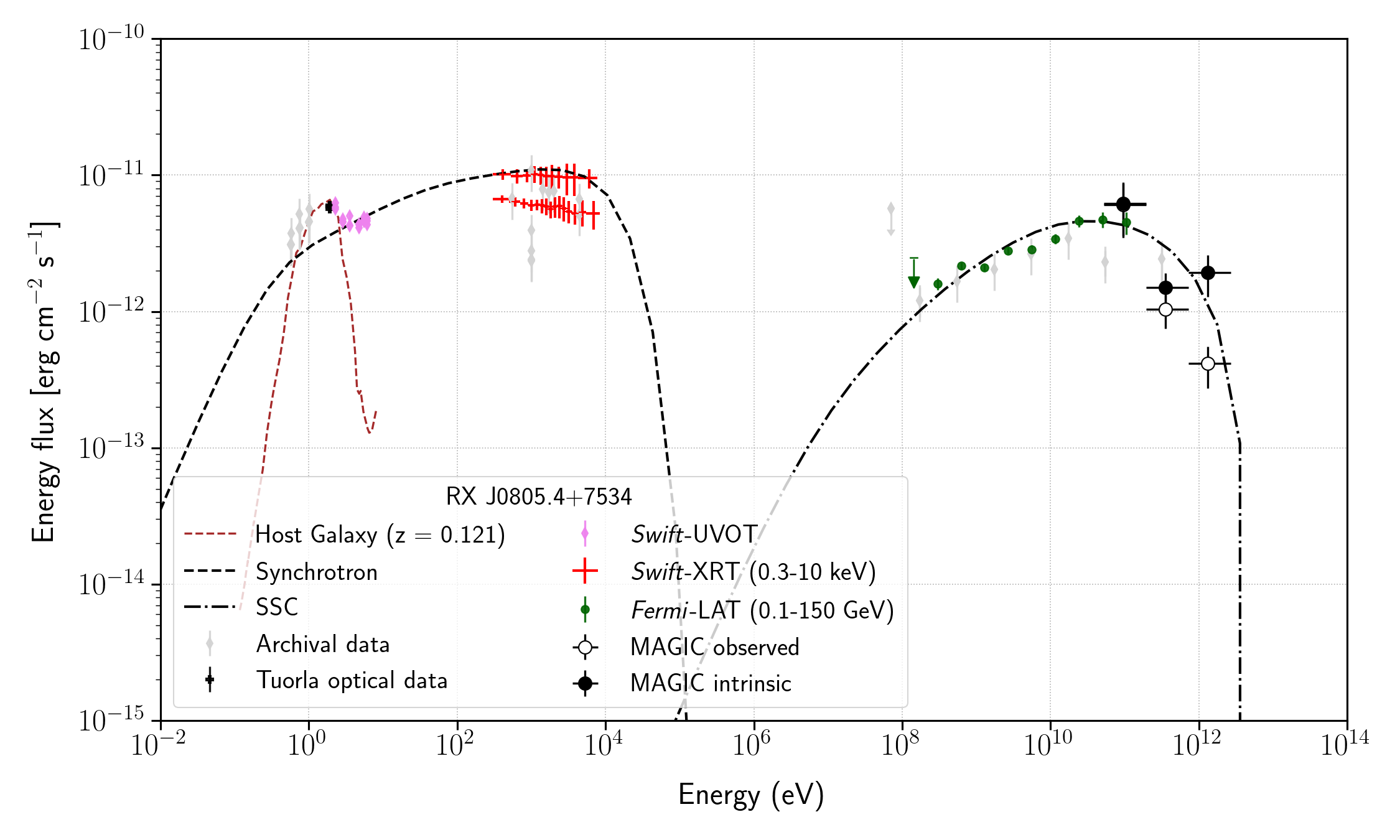}}
\subfloat[RX~J0812.0+0237.]{\includegraphics[width=0.5\textwidth]{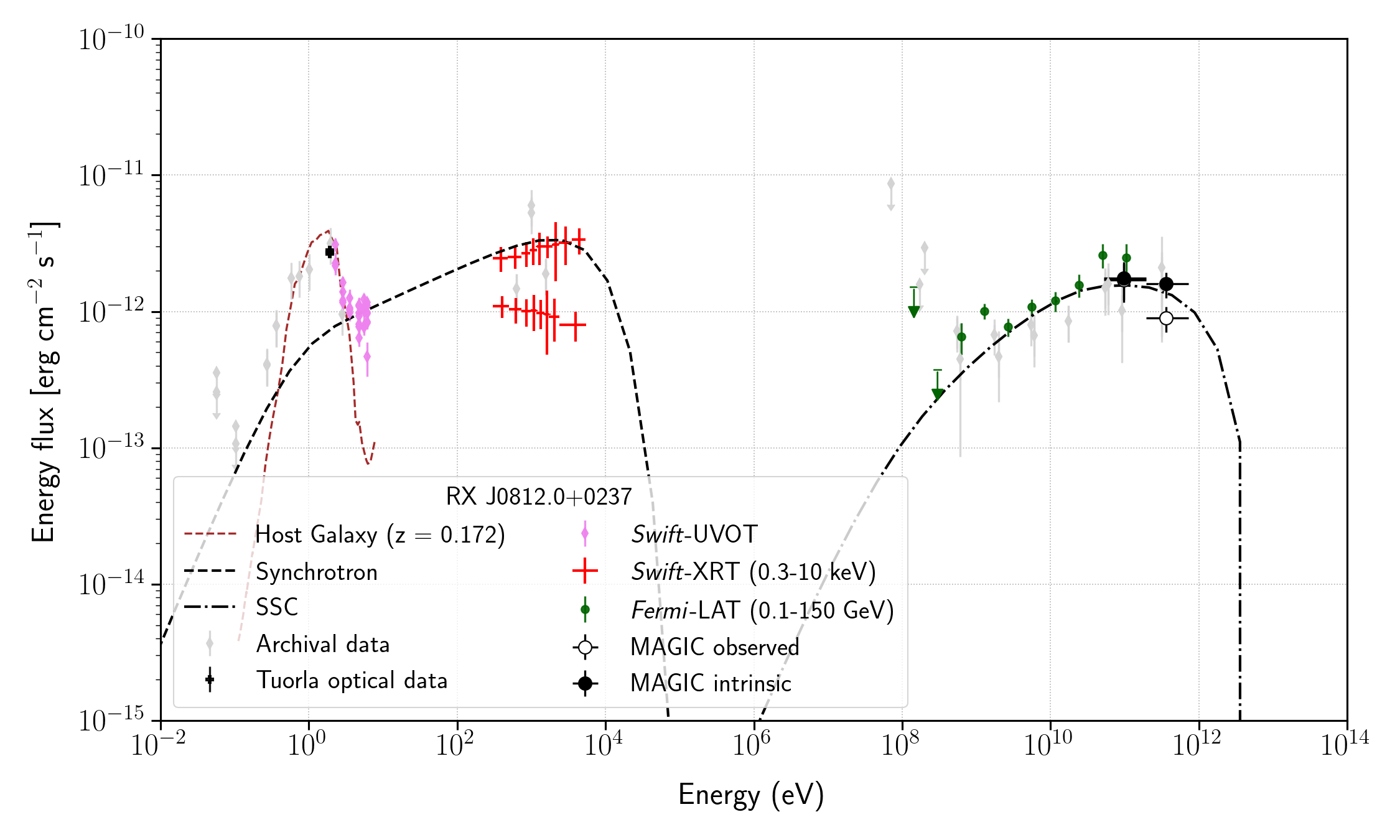}}
\end{figure*}

\begin{figure*}
\centering
\ContinuedFloat
\caption{Continues \Cref{fig:mwl_sed}.}
\subfloat[1ES~1028+511.]{\includegraphics[width=0.5\textwidth]{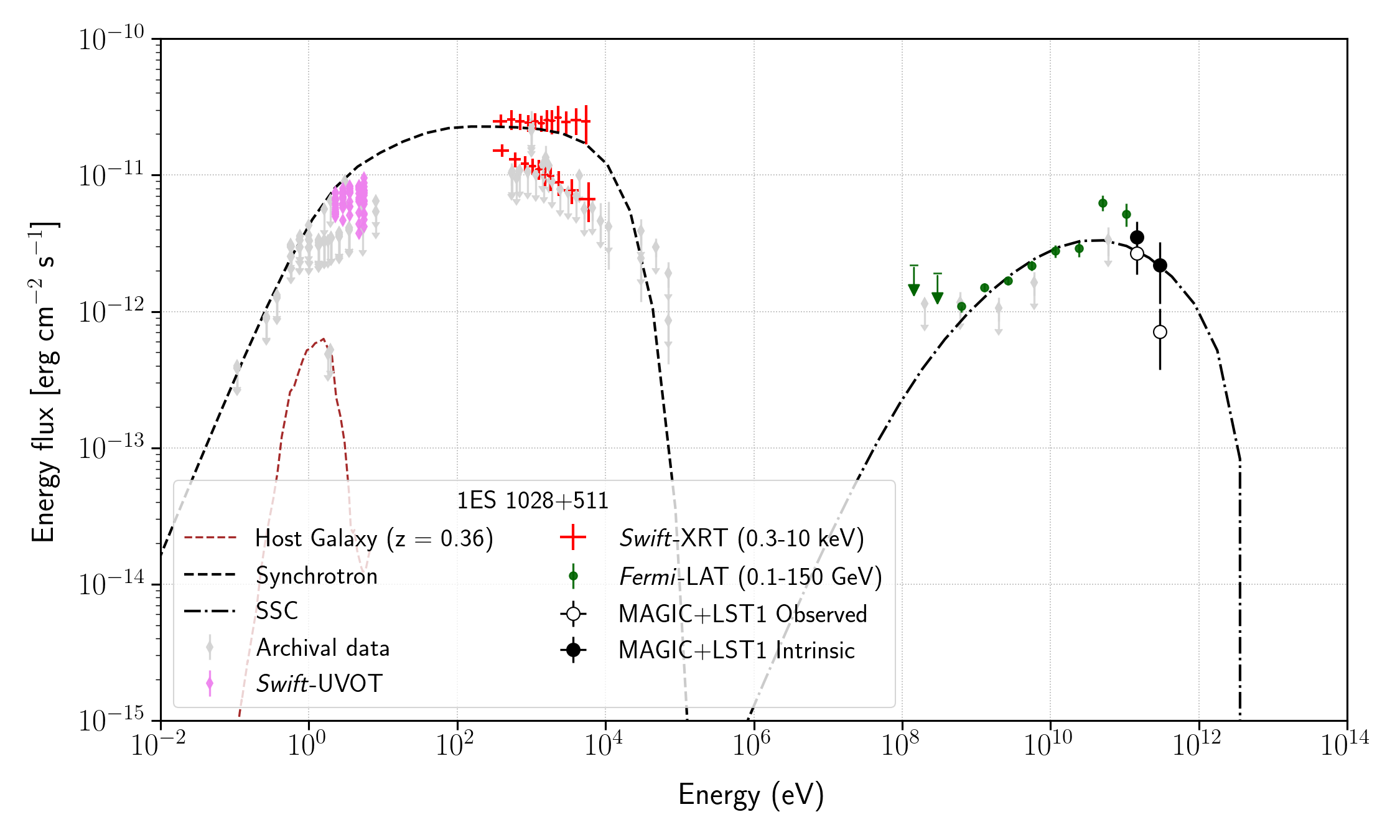}}
\subfloat[1ES~1426+428.]{\includegraphics[width=0.48\textwidth]{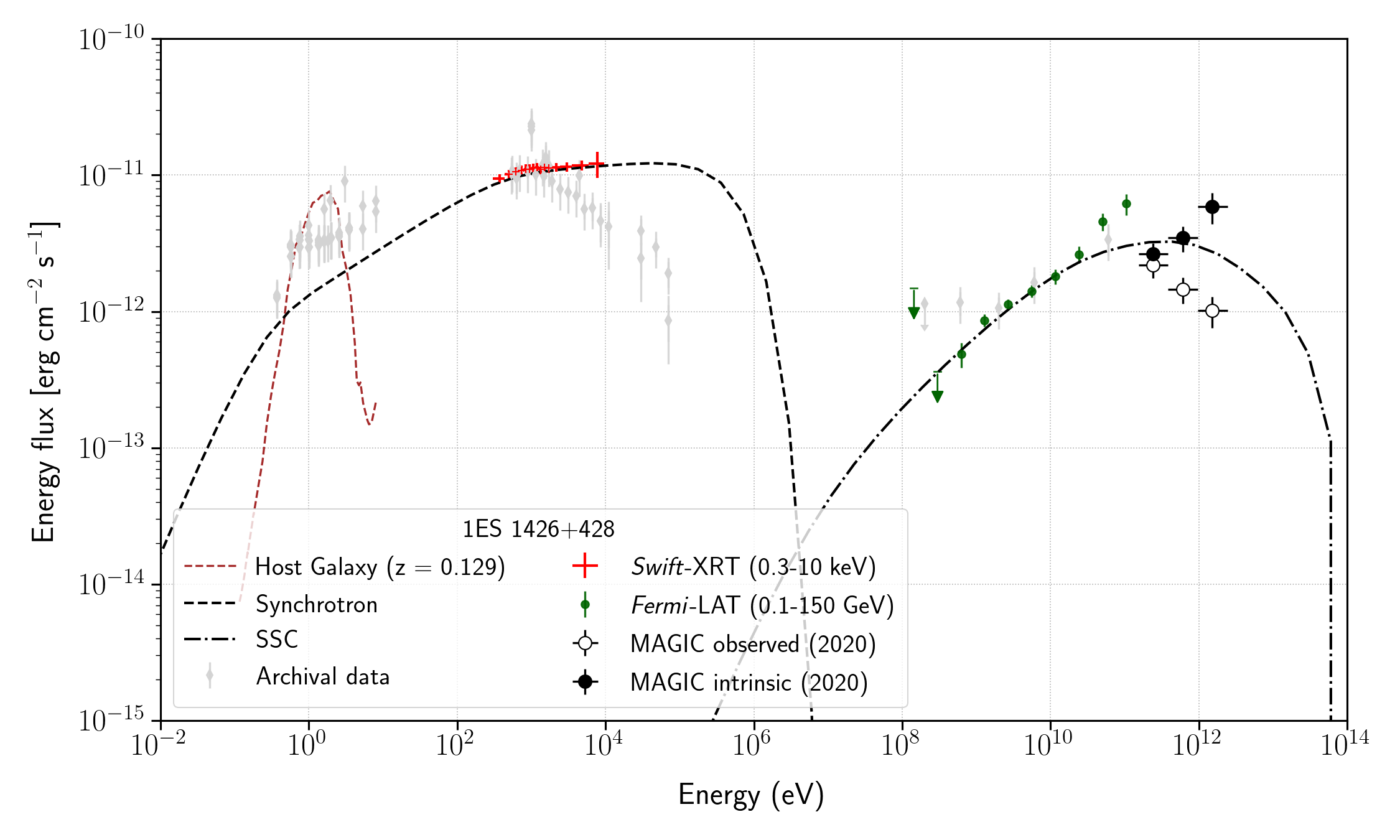}}\\
\subfloat[1ES~2037+521.]{\includegraphics[width=0.5\textwidth]{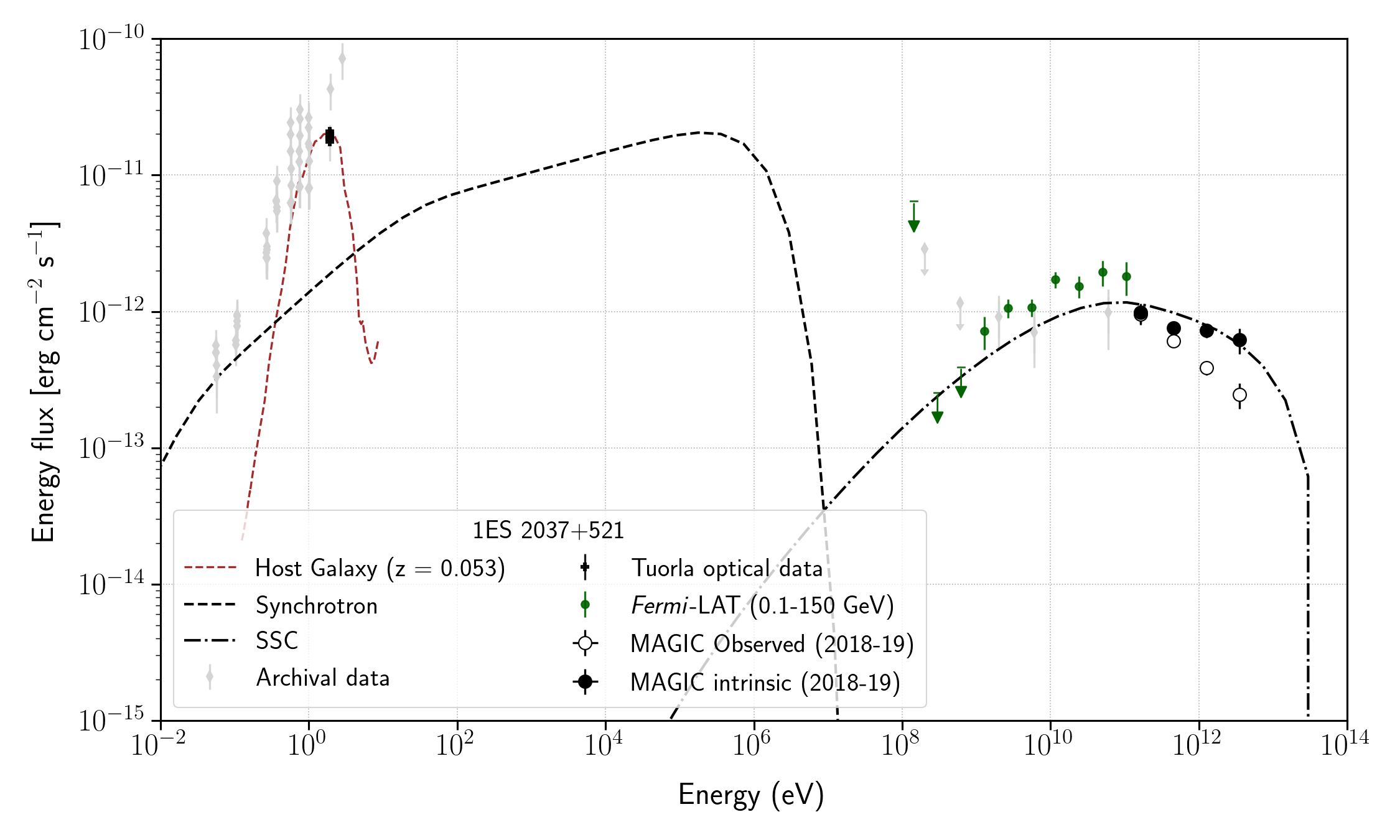}}
\end{figure*}

\section{Discussions and Conclusions}
\label{sec:discussions}
\noindent
In this paper, we presented the results of a multi-year observational campaign conducted by the MAGIC Collaboration, aimed at characterizing seven extreme high-frequency peaked BL Lac objects (extreme blazars, or EHBLs). The sources were selected as best extreme-blazar candidates to be observed at VHE gamma rays by adopting a set of complementary criteria. They were observed with the MAGIC telescopes from 2017 to 2025, for a total of 338~h of exposure, out of which 262~h of good-quality data. To enlarge the multiwavelength coverage for the modeling, the MAGIC data were complemented by  simultaneous (or quasi simultaneous, where applicable) observations from the {\it Swift}-XRT and UVOT, and with specific analyses of the \textit{Fermi}-LAT data over periods covering the MAGIC observations. For a specific source, we also report results of the joint analysis of MAGIC and LST-1 data, obtained with a set of simultaneous observations. 

This new MAGIC catalog of EHBLs reports the new detection of two sources at TeV gamma rays, namely RX~J0812.0+0237 and RX~J0805.4+7534, and the confirmation of the detection of the source 1ES~1028+511 by the MAGIC and LST-1 telescopes, after being announced in 2024 as a new TeV gamma-ray detection by the VERITAS Collaboration.
Two sources have been further monitored by MAGIC, i.e. 1ES~2037+521 and 1ES~1426+428, providing new datasets to be compared to past observations. Interestingly, although they show similar signal detections, these two sources show different results: while the former exhibits a lower state and spectrum with respect to the past, the latter shows results compatible with those found in the previous dataset. 
Additionally, this paper reports on the monitoring of the TeV candidate TXS~0637--128. The analysis of this source, which was not detected in the previous publication, now reports a hint of signal in the new dataset. Finally, this catalog also reports a nondetection for RBS~42.

The large number of sources investigated with this catalog provides key indications of the properties of extreme blazars. For example, for all of them, the high historical X-ray variability confirms the high emission activity of the tail of the particle distribution which is characterizing this energy band. On the contrary, the optical-UV band shows a more moderate activity, which is in agreement with the different cooling times of the synchrotron photons contributing at these energies. Interestingly, this low activity is also reflected on the high-energy gamma-ray band, which has shown moderate or low flux variability for all candidates and rather hard spectra. 
At VHE gamma-ray energies, these seven targets have shown different phenomenological properties.
Among the new TeV detections and the new monitored sources, most of them have shown spectra which are in agreement with the definition of soft-TeV extreme blazars. On the contrary, the source 1ES~1426+428, even if only marginally detected, displays a spectrum that is compatible with a hard-TeV extreme blazar. This result is expected and is in line with the previous findings for this source, but it emerges as a key difference with respect to the other sources of this catalog. 

The diverse behaviors of extreme blazars in this catalog clearly represent the different phenomenology of this population of sources, especially when investigated at VHE. 
Detailed broadband observations of EHBLs are essential for the discovery of new sources of this population, whose increased statistics will be key to test emission models on these sources, also depending on their activity status. The faint VHE signal of EHBLs with respect to other blazars will be faced with the improved sensitivities of the next generation of IACTs. However, the current activities of coordinated multi-frequency observations of possible candidate emitters at VHE will pave the way to the selection of best candidates for the next generation of IACTs, i.e. the CTAO, improving the possibility of future discoveries.\\

\section*{Author contribution statement}
\noindent
The MAGIC telescope system was designed and constructed by the MAGIC Collaboration. Operation, data processing, calibration, Monte Carlo simulations of the detector, and of theoretical models, and data analyses were performed by the members of the MAGIC Collaboration, who also discussed and approved the scientific results. All MAGIC collaborators contributed to the editing and comments to the final version of the manuscript. 

Luca Foffano led the project and the related observational proposals (since 2019), coordinated the data analyses -- performing several of them, as detailed below -- discussed the theoretical modeling and interpretation, and was responsible for structuring and editing the manuscript. Cornelia Arcaro performed the analysis of MAGIC data of TXS~0637-121, 1ES~1028+511, 1ES~1426+428, and 1ES~2037+521. Analysis cross-checks for the these four sources were provided by David Linder, João G. G. Paiva,  Davide Miceli, and Paolo Da Vela, respectively. Elisa Prandini, who coordinated the proposals before 2019, and Axel Arbet-Engels analysed the MAGIC data of RBS~42. João G. G. Paiva and Mireia Nievas Rosillo performed the {\it Fermi}-LAT analyses. Filippo D'Ammando performed the \textit{Swift}-XRT and UVOT analyses. Vandad F. Ramazani contributed to the project with the proposal of RX~J0805.4+7534, which was analysed by Luca Foffano and cross-checked by Cornelia Arcaro. Sweta Menon and Elisa Visentin performed the analysis of joint MAGIC + LST-1 data of 1ES~1028+511.
RX~J0812.0+0237	was analysed by Luca Foffano, Sofia Ventura, Jenni Jormanainen. 

The rest of the authors have contributed in one or several of the following ways: design, construction, maintenance and operation of the instrument(s); preparation and/or evaluation of the observation proposals; data acquisition, processing, calibration and/or reduction; production of analysis tools and/or related Monte Carlo simulations; discussion and approval of the contents of the draft.

\software{MARS} \citep{Zanin:2013oib}, 
\texttt{magic-cta-pipe}\footnote{\url{https://github.com/cta-observatory/magic-cta-pipe}}, \texttt{cta-lstchain} \citep{2022ASPC..532..357L}, \texttt{ctapipe\_io\_magic} package\footnote{\url{https://github.com/cta-observatory/ctapipe_io_magic}},  \texttt{gammapy} \citep{2023A&A...678A.157D, 2024zndo..10726484A}, \texttt{fermitools} \citep{fermitools}, \texttt{fermipy} \citep{fermipy}.
To compute the SED we made use of \texttt{agnpy}, a \texttt{python} package modeling the radiative processes of jetted AGN \citep{agnpy_paper};
version 0.4.0 of \citet{agnpy_zenodo} of the package was employed.\\


\section*{Acknowledgements}
%
%
\noindent
Part of this work is based on archival data, software, or online services provided by the Space Science Data Center – ASI. This research has made use of data and/or software provided by the High Energy Astrophysics Science Archive Research Center (HEASARC), which is a service of the Astrophysics Science Division at NASA/GSFC and the High Energy Astrophysics Division of the Smithsonian Astrophysical Observatory.

[MAGIC] We would like to thank the Instituto de Astrof\'{\i}sica de Canarias for the excellent working conditions at the Observatorio del Roque de los Muchachos in La Palma. The financial support of the German BMFTR, MPG and HGF; the Italian INFN and INAF; the Swiss National Fund SNF; the grants PID2022-136828NB-C41, PID2022-137810NB-C22, PID2022-138172NB-C41, PID2022-138172NB-C42, PID2022-138172NB-C43, PID2022-139117NB-C41, PID2022-139117NB-C42, PID2022-139117NB-C43, PID2022-139117NB-C44, CNS2023-144504 funded by the Spanish MCIN/AEI/ 10.13039/501100011033 and "ERDF A way of making Europe; the Indian Department of Atomic Energy; the Japanese ICRR, the University of Tokyo, JSPS, and MEXT; the Bulgarian Ministry of Education and Science, National RI Roadmap Project DO1-400/18.12.2020 and the Academy of Finland grant nr. 320045 is gratefully acknowledged. This work has also been supported by Centros de Excelencia ``Severo Ochoa'' y Unidades ``Mar\'{\i}a de Maeztu'' program of the Spanish MCIN/AEI/ 10.13039/501100011033 (CEX2019-000918-M, CEX2021-001131-S, CEX2024001442-S), by AST22\ 00001\_9 with funding from NextGenerationEU funds and by the CERCA institution and grants 2021SGR00426, 2021SGR00607 and 2021SGR00773 of the Generalitat de Catalunya; by the Croatian Science Foundation (HrZZ) Project IP-2022-10-4595 and the University of Rijeka Project uniri-prirod-18-48; by the Deutsche Forschungsgemeinschaft (SFB1491) and by the Lamarr-Institute for Machine Learning and Artificial Intelligence; by the Polish Ministry Of Education and Science grant No. 2021/WK/08; by the European Union (ERC, MicroStars, 101076533); and by the Brazilian MCTIC, the CNPq Productivity Grant 309053/2022-6 and FAPERJ Grants E-26/200.532/2023 and E-26/ 211.342/2021.\\

\noindent
[LST] We gratefully acknowledge financial support from the following agencies and organisations:\\
Conselho Nacional de Desenvolvimento Cient\'{\i}fico e Tecnol\'{o}gico (CNPq) Grant 309053/2022-6 and Funda\c{c}\~{a}o de Amparo \`{a} Pesquisa do Estado do Rio de Janeiro (FAPERJ) Grants E-26/200.532/2023 and E-26/211.342/2021, Funda\c{c}\~{a}o de Amparo \`{a} Pesquisa do Estado de S\~{a}o Paulo (FAPESP), Funda\c{c}\~{a}o de Apoio \`{a} Ci\^encia, Tecnologia e Inova\c{c}\~{a}o do Paran\'a - Funda\c{c}\~{a}o Arauc\'aria, Ministry of Science, Technology, Innovations and Communications (MCTIC), Brasil;
Ministry of Education and Science, National RI Roadmap Project DO1-153/28.08.2018, Bulgaria;
Croatian Science Foundation (HrZZ) Project IP-2022-10-4595, Rudjer Boskovic Institute, University of Osijek, University of Rijeka, University of Split, Faculty of Electrical Engineering, Mechanical Engineering and Naval Architecture, University of Zagreb, Faculty of Electrical Engineering and Computing, Croatia;
Ministry of Education, Youth and Sports, MEYS  LM2023047, EU/MEYS CZ.02.1.01/0.0/0.0/16\ 013/ 0001403,  CZ.02.1.01/0.0/0.0/18\ 046/ 0016007, CZ.02.1.01/0.0/0.0/16\ 019/ 0000754, CZ.02.01.01/ 00/22\ 008/ 0004632 and CZ.02.01.01/ 00/23\ 015/ 0008197 Czech Republic;
CNRS-IN2P3, the French Programme d’investissements d’avenir and the Enigmass Labex, 
This work has been done thanks to the facilities offered by the Univ. Savoie Mont Blanc - CNRS/IN2P3 MUST computing center, France;
Max Planck Society, German Bundesministerium f{\"u}r Forschung, Technologie und Raumfahrt (Verbundforschung / ErUM), the Deutsche Forschungsgemeinschaft (SFB 1491) and the Lamarr-Institute for Machine Learning and Artificial Intelligence, Germany;
Istituto Nazionale di Astrofisica (INAF), Istituto Nazionale di Fisica Nucleare (INFN), Italian Ministry for University and Research (MUR), and the financial support from the European Union -- Next Generation EU under the project IR0000012 - CTA+ (CUP C53C22000430006), announcement N.3264 on 28/12/2021: ``Rafforzamento e creazione di IR nell’ambito del Piano Nazionale di Ripresa e Resilienza (PNRR)'';
ICRR, University of Tokyo, JSPS, MEXT, Japan;
JST SPRING - JPMJSP2108;
Narodowe Centrum Nauki, grant number 2023/50/A/ST9/ 00254, Poland;
The Spanish groups acknowledge the Spanish Ministry of Science and Innovation and the Spanish Research State Agency (AEI) through the government budget lines
PGE2022/ 28.06.000X.711.04,
28.06.000X.411.01 and 28.06.000X.711.04 of PGE 2023, 2024 and 2025,
and grants PID2019-104114RB-C31,  PID2019-107847RB-C44, PID2019-105510GB-C31, PID2019-104114RB-C33, PID2019-107847RB-C43, PID2019-107847RB-C42, PID2019-107988GB-C22, PID2021-124581OB-I00, PID2021-125331NB-I00, PID2022-136828NB-C41, PID2022-137810NB-C22, PID2022-138172NB-C41, PID2022-138172NB-C42, PID2022-138172NB-C43, PID2022-139117NB-C41, PID2022-139117NB-C42, PID2022-139117NB-C43, PID2022-139117NB-C44, PID2022-136828NB-C42, PID2024-155316NB-I00, PDC2023-145839-I00 funded by the Spanish MCIN/AEI/10.13039/ 501100011033 and by ERDF/EU and NextGenerationEU PRTR; CSIC PIE 202350E189; the "Centro de Excelencia Severo Ochoa" program through grants no. CEX2020-001007-S, CEX2021-001131-S, CEX2024-001442-S; the "Unidad de Excelencia Mar\'ia de Maeztu" program through grants no. CEX2019-000918-M, CEX2020-001058-M; the "Ram\'on y Cajal" program through grants RYC2021-032991-I  funded by MICIN/ AEI/10.13039/ 501100011033 and the European Union “NextGenerationEU”/PRTR and RYC2020-028639-I; the "Juan de la Cierva-Incorporaci\'on" program through grant no. IJC2019-040315-I and "Juan de la Cierva-formaci\'on"' through grant JDC2022-049705-I; the “Viera y Clavijo” postdoctoral program of Universidad de La Laguna, funded by the Agencia Canaria de Investigaci\'on, Innovaci\'on y Sociedad de la Informaci\'on. They also acknowledge the "Atracci\'on de Talento" program of Comunidad de Madrid through grant no. 2019-T2/TIC-12900; “MAD4SPACE: Desarrollo de tecnolog\'ias habilitadoras para estudios del espacio en la Comunidad de Madrid" (TEC-2024/TEC-182) project, Doctorado Industrial (IND2024/TIC34250) and Ayudas para la contrataci\'on de personal investigador predoctoral en formación (PIPF-2023/TEC-29694) funded by Comunidad de Madrid; the La Caixa Banking Foundation, grant no. LCF/BQ/PI21/11830030; Junta de Andaluc\'ia under Plan Complementario de I+D+I (Ref. AST22\ 0001) and Plan Andaluz de Investigaci\'on, Desarrollo e Innovaci\'on as research group FQM-322; Project ref. AST22\ 00001\_9 with funding from NextGenerationEU funds; the “Ministerio de Ciencia, Innovaci\'on y Universidades”  and its “Plan de Recuperaci\'on, Transformaci\'on y Resiliencia”; “Consejer\'ia de Universidad, Investigaci\'on e Innovaci\'on” of the regional government of Andaluc\'ia and “Consejo Superior de Investigaciones Cient\'ificas”, Grant CNS2023-144504 funded by MICIU/AEI/10.13039/501100011033 and by the European Union NextGenerationEU/PRTR,  the European Union's Recovery and Resilience Facility-Next Generation, in the framework of the General Invitation of the Spanish Government's public business entity Red.es to participate in talent attraction and retention programmes within Investment 4 of Component 19 of the Recovery, Transformation and Resilience Plan; Junta de Andaluc\'{\i}a under Plan Complementario de I+D+I (Ref. AST22\ 00001), Plan Andaluz de Investigaci\'on, Desarrollo e Innovación (Ref. FQM-322). ``Programa Operativo de Crecimiento Inteligente" FEDER 2014-2020 (Ref.~ESFRI-2017-IAC-12), Ministerio de Ciencia e Innovaci\'on, 15\% co-financed by Consejer\'ia de Econom\'ia, Industria, Comercio y Conocimiento del Gobierno de Canarias; the "CERCA" program and the grants 2021SGR00426 and 2021SGR00679, all funded by the Generalitat de Catalunya; and the European Union's NextGenerationEU (PRTR-C17.I1). This work is funded/Co-funded by the European Union (ERC, MicroStars, 101076533). This research used the computing and storage resources provided by the Port d'Informaci\'o Cient\'ifica (PIC) data center.
State Secretariat for Education, Research and Innovation (SERI) and Swiss National Science Foundation (SNSF), Switzerland;
The research leading to these results has received funding from the European Union's Seventh Framework Programme (FP7/2007-2013) under grant agreements No~262053 and No~317446;
This project is receiving funding from the European Union's Horizon 2020 research and innovation programs under agreement No~676134;
ESCAPE - The European Science Cluster of Astronomy \& Particle Physics ESFRI Research Infrastructures has received funding from the European Union’s Horizon 2020 research and innovation programme under Grant Agreement no. 824064.


\bibliography{biblio_2nd_MAGIC_EHBL_catalog}


\appendix
\section{\label{app:magic}MAGIC \lf{observations and} data analysis details}

\noindent
MAGIC is a system of two imaging atmospheric Cherenkov telescopes (IACTs) designed to detect the UV-optical Cherenkov light produced when a gamma ray interacts with the atmosphere, resulting in a cascade of superluminal, charged particles \citep{magicperf_1:2015}. Located on the Canary Island of La Palma at an altitude of 2200 m, each telescope has 17-m-diameter reflective surface, allowing MAGIC to achieve an energy threshold as low as 50 GeV in standard-trigger mode. For point-like sources, the integral sensitivity above 220 GeV is ($0.83 \pm 0.03$)\% of the Crab Nebula flux after 50 hours of observation, assuming a Crab Nebula-like spectrum. At these energies, the angular resolution is below 0.07 degrees, with an energy resolution of 16\%. Details on the performance and data analysis methods for this instrument are fully covered in \citet{magicperf_2:2015}. 

The energy threshold is particularly important in the study of EHBLs at VHE. In the MAGIC analysis, it is estimated from the peak of the simulated Monte Carlo gamma-ray events, weighting their spectrum by the source spectrum. 
Key factors affecting the energy threshold include the zenith angle of observations and background light conditions. Medium and high zenith angles (above 35° and 50°, respectively) result in an increased energy threshold as particle showers pass through a thicker atmospheric layer. This increased threshold is accompanied by higher sensitivity at the upper energy range, thanks to an enlarged effective area \citep{magicperf_2:2015}. High background light from the Moon also influences the energy threshold; however, {a specific cleaning procedure can be applied in order to analyze these data} \citep{magicperf_moon:2017}. 

\lf{All observations were performed in the so-called wobble pointing mode,  with the source position offset by 0.4 degrees}. Data were analyzed using the MAGIC Analysis and Reconstruction Software package \citep[MARS,][]{Zanin:2013oib}, adapted for stereoscopic observations \citep{2009arXiv0907.0943M}. A key parameter, $\theta^2$, was used to identify significant VHE gamma-ray excesses, defined as the squared angular distance between the reconstructed shower direction and the source position in the camera.  The VHE gamma-ray signal is typically concentrated at low $\theta^2$ values in the ``On'' region, with the cosmic-ray background estimated from three ``Off'' regions located at the same radial distance from the source position at 90, 180, and 270 degrees.\\


\section{LST-1 observations and data analysis}
\label{sec:LST_data_analysis_details}

\subsection{Introduction}
\noindent
The new generation of Cherenkov telescopes for the investigation of the VHE gamma-ray sky is represented by the Cherenkov Telescopes Array Observatory - CTAO\footnote{\url{https://www.ctao.org/}}, which is currently under construction \citep{2013APh....43....3A}. Thanks to an improvement of the  sensitivity of the current generation of Cherenkov Telescopes of more than 1 order of magnitude at 1~TeV, it will {advance} the scientific investigation of this energy range.
The observatory will include three types of telescopes, with different designs and key energy sensitivities, in such a way to cover a wide energy range with the full-array sensitivity between 20 GeV and 300 TeV.
Among them, {the LSTs represent} the largest telescope type. With their 23-m-diameter mirror dish, they have a large light collection area aimed at covering the low-energy band from tens to hundreds of GeV \citep{LSTPerformance}.

\mbox{LST-1} is the first of four LSTs that will constitute {the CTAO's Northern Hemisphere} \citep[CTAO-North;][]{2019APh...111...35A}, which is located at the Observatorio del Roque de los Muchachos on the Canary island of La Palma (Spain). Since October 2018, when its construction was completed, LST-1 started a  commissioning and validation period. During this period, the telescope has also been conducting scientific observations.

\subsection{Joint analysis of simultaneous MAGIC and LST-1 data}
\noindent
Joint simultaneous observations of 1ES~1028+511 with MAGIC and LST-1 -- with the same wobble pointing mode and an offset of 0.4 degrees -- were performed in a period between April 29, 2022 and February 11, 2024, for a total of 15.6~h before quality cuts, as detailed in \Cref{tab:obs_table}. 
These simultaneous observations between the two telescopes provide increased collection area and improved background rejection, resulting in an improved sensitivity with respect to standalone observations. 
Details on the performance of the MAGIC + LST joint observations and on the specific data analysis chain are reported in \citet{MAGIC_LST_joint_performance}.

In this work, we performed a joint analysis of MAGIC and LST-1 simultaneous data. During standalone operations, both MAGIC and LST-1 use independent analysis chains: \texttt{MARS} for the former, and \texttt{cta-lstchain} for the latter. 
The joint analysis pipeline {matches event timestamps} from LST‑1 and both MAGIC telescopes, reconstructing stereoscopic images for coincident air showers.
The MAGIC data at raw level are stored in a custom binary format and are not compatible with the LST-1 raw data, which consist of pixel-wise waveforms and auxiliary information. 
Initially, signal extraction from individual pixel waveforms and  proper calibration are performed for each instrument with the specific software. The calibrated MAGIC data, once converted into \texttt{HDF5} format using the dedicated \texttt{ctapipe\_io\_magic} package\footnote{\url{https://github.com/cta-observatory/ctapipe_io_magic}}, are then compatible with the LST-1 DL1 processed data. The following steps of the analysis chain are performed with the \texttt{magic-cta-pipe} package using \texttt{lst-chain} and \texttt{ctapipe} methods \citep{karl_kosack_2024_13757224}. 

MAGIC + LST-1 simultaneous data were selected by considering only dark or low-moonlight observing conditions (observations with the Moon below the horizon or maximum diffuse night-sky-background level below 2.3 photoelectrons per second) and with good atmospheric conditions (indicated by stable rates, atmospheric transmission at 9\,km above 80\%, and cloudiness below 20\%). We also applied a cut on the minimal rate of differential cosmic rays as a function of intensity of 1.5 at 422 photoelectrons. Furthermore, we excluded runs affected by data acquisition issues. 
The final amount of data after such a quality selection results in a total of 6.0~h of observations joint with MAGIC, detailed in \Cref{tab:obs_table}.

The high-level scientific products such as the event $\theta^2$ distribution plots and joint spectrum were obtained adopting the open-source software package \texttt{Gammapy},{ version 1.3.}
The analysis was performed by adopting standard control sky regions (OFF regions, used to subtract the background) located at angular rotation offsets of 90\degr, 180\degr\, and 270\degr\, where 0\degr\ is coincident with position of 1ES~1028+511. 

The energy threshold of the joint analysis is determined on the basis of the specific Instrument Response Functions (IRFs), and with a minimum effective area of 10\% of its maximum value within the region where the spectral analysis shows significant excess. To estimate the value of the energy threshold, we adopt the peak of the simulated Monte Carlo gamma-ray events, weighting their spectrum by the source spectrum of $\Gamma \sim 3.5$. The final value, considering its dependence on the zenith angle of the observations, is estimated as $E\sim 80 \,\rm{GeV}$. However, since in our analysis we explore the results of the joint dataset together with MAGIC-only data, we apply the latter energy threshold on the flux estimation.
Given the relatively faint signal detection, we computed the integral photon flux using a power-law spectral shape from the best-fit model, which results in $4.5 \pm 1.1 \times 10^{-12}$ cm$^{-2}$ s$^{-1}$ {above 150~GeV}.\\

\begin{table}
    \begin{center}
    \caption{Observation campaign with \mbox{LST-1} of 1ES~1028+511. For each observation day, we list the starting date, zenith range, {the  observation time before and after (\textit{effective} time) any data quality selection.}}
    \label{tab:obs_table}
        \begin{tabular}{ccccc}
        \hline \hline
        Start date &  Zenith range &  Obs. time & Eff. time  \\
          & [deg] & [h] & [h] \\
        \hline
2022/04/29	&	22	-	26	&	0.9	&  0.8  \\
2022/05/22	&	35	-	44	&	0.7	&  0.3  \\
2023/03/19	&	22	-	25	&	0.7	&  0.6  \\
2023/03/20	&	24	-	37	&	1.9	&  1.8 \\
2023/03/21	&	23	-	48	&	3.5	&  2.0  \\
2023/03/22	&	23	-	66	&	0.2	&    \\
2023/03/23	&	22	-	40	&	2.2	&    \\
2023/03/24	&	25	-	38	&	1.5	&    \\
2023/03/25	&	26	-	37	&	1.4	&    \\
2023/03/26	&	33	-	43	&	1.0	&    \\
2024/02/11	&	22	-	31	&	1.6	&  0.5\\
\hline							
Total	&	22	-	66	&	15.6	& 6.0 \\
        \hline
        \end{tabular}
    \end{center}
\end{table}


\section{{\it Fermi}-LAT data analysis details}
\label{app:fermi_analysis}
\noindent
The {\em Fermi} LAT \citep{2009ApJ...697.1071A} is a pair conversion telescope consisting of a $4\times4$ array of silicon strip trackers and tungsten converters and a Cesium Iodine (CsI) based calorimeter. The instrument is fully covered by a segmented anti-coincidence shield which provides a highly efficient vetoing against charged particle background events. The LAT is sensitive to gamma rays from $20\,{\rm MeV}$ to more than $300\,{\rm GeV}$ and normally operates in survey mode, covering the whole sky every three hours and providing an instantaneous field of view (FOV) of $2.4\,{\rm sr}$. 

The LAT data was extracted from the weekly data files available at the FSSC data center\footnote{\protect\url{https://fermi.gsfc.nasa.gov/ssc/data/access/}}.  For each data sample, only Pass 8 source-class photons detected within $15^\circ$ of the nominal position of the analyzed source were considered. Only events whose reconstructed energy lay between $300 \,{\rm MeV}$ and $150\,{\rm GeV}$ were selected. The relatively high-energy threshold was set to simplify the analysis and remove contamination from secondary sources, which could lead to biased results if the minimum-range limit was considered below 300 MeV. Following the event selection recommendations from Cicerone\footnote{\protect\url{https://fermi.gsfc.nasa.gov/ssc/data/analysis/documentation/Cicerone/}}, only good data ({\tt (DATA\_QUAL>0)\&\&(LAT\_CONFIG==1)}) with zenith distance lower than $90^\circ$ were included.

For each data sample, the data was analyzed using \texttt{Fermitools} v2.2.0 and \texttt{Fermipy}, version 1.2. The period for which we decided to obtain the results was between 2008-08-04 15:59:59 (MJD~54682.6) and 2024-12-25 23:59:59 (MJD~60670.0). 
According to the main website, there were technical issues with the \textit{Fermi}-LAT computing cluster, resulting in some data gaps after 2024-12-25. For this reason, we decided to conclude the analysis on this day and not on 2024-12-31.
A summed binned likelihood analysis approach was followed with event type 3 (related to the detection of front and back layers of the tracker), event class 128 (classified as Source type events), and 3 bins per energy decade, using the IRFs P8R3\_SOURCE\_V3. All 4FGL sources within the region of interest (ROI) were included in the model, along with Galactic and isotropic models using the {\tt gll\_iem\_v07.fits} and {\tt iso\_P8R3\_SOURCE\_V3\_v01.txt} files, respectively. The spectra of the sources were selected such to maximize the value of the likelihood while being physically sound, following the same method described in \citet{2019MNRAS.486.4233A}. All sources were modeled with  spectral shapes attenuated by the EBL using the template from \citet{Dominguez:2010bv}. For each analysis, the spectral parameters of all sources that are significantly detected within a radius of $3^\circ$ around the ROI were left free in the fit in order to account for their possible variability. The parameters of the remaining sources were fixed to the published 4FGL values. The normalization of the diffuse components was left free. All results are reported in \Cref{tab:spectraLATandXRT}.\\

\begin{table*}
	\centering
		\caption{Main spectral parameters from the \textit{Fermi}-LAT 4FGL \citep{fermi4fgl} and 3FHL catalogs \citep{fermi3FHL}. Columns from \textit{left} to \textit{right}: common source name; 4FGL source name, HE gamma-ray photon flux in the range of 1-100 GeV, spectral index for the power-law fit within 100 MeV-100 GeV, detection significance, and variability index reported in the 4FGL catalog; photon index for the power-law fit $>$50 GeV reported in the 3FHL catalog.}
 		\label{tab:fermi_lat_catalog_info}
 		\begin{tabular}{lcccccc} 
 			\toprule
 			\multirow{2}{*}{Source} &\multirow{2}{*}{4FGL name}  & Flux$_{\mathrm{4FGL}}$ &\multirow{2}{*}{$\Gamma _{\mathrm{4FGL}}$} & Signif. & Var. Index    &  $\Gamma _{\mathrm{3FHL}}$    \\
             & &  $\times10^{-10}$ [ph\,cm$^{-2}$\,s$^{-1}$] &  & [$\sigma$] &  & $E > 10$\,GeV     \\
 			\hline
RBS~42 & 4FGL J0018.4+2946 & $2.59 \pm 0.33$ & $1.74 \pm 0.07$ & 15.6 & 13.7 & 	$1.98\pm0.29 $\\ \hline
TXS~0637--128 & 4FGL J0640.0-1253	 & $4.70 \pm 0.55$ & $1.74 \pm 0.07$ & 18.0 &7.5&$ 1.63	\pm0.19$ \\ \hline
RX~J0805.4+7534 & 4FGL J0805.4+7534 & $14.53 \pm 0.47$ & $1.84 \pm 0.02$ & 62.3 &42.0 &	$2.28\pm0.16$\\ \hline
RX~J0812.0+0237 & 4FGL J0812.0+0237 & $4.82 \pm 0.45$ & $1.88 \pm 0.07$ & 20.4 & 8.4 & $1.96\pm0.25$ \\ \hline
1ES~1028+511 & 4FGL J1031.3+5053 & $9.70\pm 0.43$ & $1.73 \pm 0.03$ & 47.2 &15.9& 	$1.94\pm0.15$ \\ \hline
1ES~1426+428 & 4FGL J1428.5+4240 & $5.77 \pm 0.35$ & $1.63\pm0.05$ & 35.5 & 23.0 & 	$1.91\pm0.14$ \\ \hline
1ES~2037+521 & 4FGL J2039.5+5218 & $4.54 \pm 0.53$ & $1.84 \pm 0.07$ & 16.1 &7.1& 	$2.11\pm0.28$ \\ \hline
\toprule
	 	\end{tabular}
\end{table*}


\section{{\it Swift} data analysis details}

\subsection{XRT data analysis}
\label{app:xrt_analysis} 

\noindent
All XRT observations were performed in photon counting (PC) mode, depending on the brightness of the source. The XRT spectra were generated with the \textit{Swift}-XRT data products generator tool at the UK Swift Science Data Centre\footnote{\url{http://www.swift.ac.uk/user\_objects}} \citep[for details see][]{2009MNRAS.397.1177E}. Spectra having count rates higher than 0.5 counts s$^{-1}$ may be affected by pile-up. To correct for this effect, the central region of the image was excluded, and the source image was obtained from an annular extraction region with an inner radius depending on the level of pile-up \citep[see e.g.,][]{moretti05}.

The X-ray spectra in the 0.3--10 keV energy range were fitted with an absorbed power-law model using the photoelectric absorption model \texttt{tbabs} \citep{wilms00} with a HI column density consistent with the Galactic value in the direction of the sources as reported in \citet{bekhti06}. The spectral uncertainties also account for systematic effects arising from the correction for HI absorption. A non-negligible amount of spectra show low number of counts (i.e. $<$ 200), resulting in a low number of spectral bins with sufficient counts to be well-approximated by a Gaussian distribution, an assumption on which the $\chi^{2}$ relies. To maintain the homogeneity in the analysis, all spectra were grouped using the task \texttt{grppha} to have at least one count per bin and the fit was performed with the Cash statistics \citep{cash79}. We used the spectral redistribution matrices in the Calibration database maintained by HEASARC. The X-ray spectral analysis was performed using the {\texttt XSPEC 12.13.1} software package \citep{arnaud96}. The  fit results are reported in \Cref{tab:spectraLATandXRT}. \\

\subsection{UVOT data analysis}
\label{app:uvot_analysis} 
\noindent
During the \textit{Swift} pointings, the UVOT instrument observed the sources in its optical ($v$, $b$ and $u$) and UV ($w1$, $m2$ and $w2$) photometric bands \citep{poole08,breeveld10}. UVOT data in all filters were analyzed with the \texttt{uvotimsum} and \texttt{uvotsource} tasks included in the HEASoft package (v6.33.1) and the 20240201 CALDB-UVOTA release. Source counts were extracted from a circular region of 5 arcsec radius centered on the source, while background counts were derived from a circular region with 20 arcsec radius in a nearby source-free region.
The UVOT magnitudes were corrected for Galactic extinction using the E(B--V) value from \citet{schlafly11} and the extinction laws from \citet{cardelli89} and converted to flux densities using the conversion factors from \citet{breeveld10}.

\begin{table}
\centering
\caption{Parameter values derived from the SSC leptonic model used in this study. We report the bulk Lorentz factor $\Gamma$, the magnetic field $B$, the size of the emitting region $R$, and the electron spectral normalization $k_e$. The electron distribution is assumed to have an index of $p_1$ between $\gamma_{\text{min}}$ and $\gamma_{\text{br}}$, and an index of $p_2$ up to the maximum $\gamma_{\text{max}}$. }
\label{tab:modeling}
    \begin{tabular}{lcccccccccccccc}
        \hline\hline
    Source	   & 
    $\Gamma$  &  $B$ &  $R$ & $k_{\rm e}$  & $\gamma_{\rm min}$& $\gamma_{\rm br}$ & $\gamma_{\rm max}$ & $p_1$ & $p_2$  \\
     &  &  [G] & [$\mbox{cm}$] & [cm$^{-3}$] & & &  &  &  \\
    \hline
    \lf{TXS~0637--128} & $15$ & 0.12 &$2 \times 10^{16}$&  $1 \times 10^{-7}$ & 7000 &$ 1 \times 10^{4}$ & $6 \times 10^{5}$ & 1.6 & 2.4  \\
    RX~J0805.4+7534& $15$ & 0.12 &$2 \times 10^{16}$&  $4 \times 10^{-7}$ & 3000 &$ 4 \times 10^{4}$ & $7 \times 10^{5}$ & 2.4 & 2.8  \\
    RX~J0812.0+0237 & 15 & 0.06 &$2 \times 10^{16}$&  $2 \times 10^{-6}$ & 6000 &$ 1 \times 10^{4}$ & $8 \times 10^{5}$ & 1.8 & 2.5  \\
    1ES~1028+511 & $30$ & 0.2 &$7 \times 10^{15}$&  $5 \times 10^{-5}$ & 2000 &$ 1 \times 10^{4}$ & $3 \times 10^{5}$ & 2.4 & 2.8  \\
    1ES~1426+428 & $27$ & 0.15 &$5 \times 10^{16}$&  $1.2 \times 10^{-7}$ & 5000 &$ 3 \times 10^{4}$ & $5 \times 10^{5}$ & 2.3 & 3.0\\
    1ES~2037+521 & $12$ & 0.11 &$2 \times 10^{16}$&  $1.5 \times 10^{-7}$ & 1000 &$ 4 \times 10^{4}$ & $8 \times 10^{6}$ & 2.0 & 2.7\\
    \hline\hline
    \end{tabular}
\end{table}

\label{lastpage}
\end{document}

%% file: authorlist.tex
\author{K.~Abe}
\affiliation{Department of Physics, Tokai University, 4-1-1, Kita-Kaname, Hiratsuka, Kanagawa 259-1292, Japan}
\author[0000-0001-7250-3596]{S.~Abe}
\affiliation{Institute for Cosmic Ray Research, University of Tokyo, 5-1-5, Kashiwa-no-ha, Kashiwa, Chiba 277-8582, Japan}
\author[0000-0001-8215-4377]{J.~Abhir}
\affiliation{ETH Z\"urich, CH-8093 Z\"urich, Switzerland}
\author[0009-0005-5239-7905]{A.~Abhishek}
\affiliation{INFN and Università degli Studi di Siena, Dipartimento di Scienze Fisiche, della Terra e dell'Ambiente (DSFTA), Sezione di Fisica, Via Roma 56, 53100 Siena, Italy}
\author[0000-0001-8307-2007]{V.~A.~Acciari}
\affiliation{Institut de Fisica d'Altes Energies (IFAE), The Barcelona Institute of Science and Technology, Campus UAB, 08193 Bellaterra (Barcelona), Spain}
\author[0000-0002-6606-2816]{F.~Acero}
\affiliation{Université Paris-Saclay, Université Paris Cité, CEA, CNRS, AIM, F-91191 Gif-sur-Yvette Cedex, France}
\affiliation{FSLAC IRL 2009, CNRS/IAC, La Laguna, Tenerife, Spain}
\author[0000-0001-8816-4920]{A.~Aguasca-Cabot}
\affiliation{Departament de Física Quàntica i Astrofísica, Institut de Ciències del Cosmos, Universitat de Barcelona, IEEC-UB, Martí i Franquès, 1, 08028, Barcelona, Spain}
\author[0000-0002-3777-6182]{I.~Agudo}
\affiliation{Instituto de Astrofísica de Andalucía-CSIC, Glorieta de la Astronomía s/n, 18008, Granada, Spain}
\author{C.~Alispach}
\affiliation{Department of Astronomy, University of Geneva, Chemin d'Ecogia 16, CH-1290 Versoix, Switzerland}
\author{D.~Ambrosino}
\affiliation{INFN Sezione di Napoli, Via Cintia, ed. G, 80126 Napoli, Italy}
% ...

\collaboration{for the MAGIC and CTAO-LST Collaborations, and}

\author[0000-0001-7618-7527]{F.~D'Ammando}
\affiliation{INAF - Istituto di Radioastronomia, Via Piero Gobetti 93/3, 40129 Bologna, Italy}

\author[0000-0001-8991-7744]{V.~Fallah Ramazani}
\affiliation{Finnish MAGIC Group: Finnish Centre for Astronomy with ESO, Department of Physics and Astronomy, University of Turku, FI-20014 Turku, Finland}
\author{D.~Linder}
\affiliation{ETH Z\"urich, CH-8093 Z\"urich, Switzerland}

\author[0000-0002-7708-6362]{F.~Wierda}
\affiliation{Finnish MAGIC Group: Finnish Centre for Astronomy with ESO, Department of Physics and Astronomy, University of Turku, FI-20014 Turku, Finland}